\documentclass[sigconf]{acmart}

\PassOptionsToPackage{protrusion, expansion}{microtype} 

\usepackage{amsmath, amsthm}
    
\usepackage{graphicx, xcolor} 
\usepackage{hyperref, url}
\usepackage{adjustbox}
\usepackage{booktabs, multirow}
\usepackage{caption}
\usepackage{subcaption}
\usepackage{enumitem}
\usepackage{lipsum}
\usepackage{setspace}
\usepackage{algorithm, algpseudocode}
    \definecolor{BLUE}{rgb}{.0, .2, .6}
    \definecolor{BLUEalt}{HTML}{1e50a2}
    \definecolor{RED}{HTML}{c9171e}
    \algrenewcommand{\alglinenumber}[1]{{\scriptsize\bfseries\ttfamily\color{RED}#1}}

\usepackage{footnote}
\makesavenoteenv{tabular}
\makesavenoteenv{table}

\setcopyright{none}
\renewcommand\footnotetextcopyrightpermission[1]{}
\usepackage[zerostyle=d]{newtxtt}

\begin{document}

\title{A Novel Memory-Efficient Deep Learning Training Framework via Error-Bounded Lossy Compression}

\newcommand{\AFFIL}[4]{
    \affiliation{
        \institution{\small #1}
        \city{#2}\state{#3}\country{#4}
    }
    }

\author{Sian Jin}{\AFFIL{Washington State University}{Pullman}{WA}{USA}}
\email{sian.jin@wsu.edu}

\author{Guanpeng Li}{\AFFIL{University of Iowa}{Iowa City}{IA}{USA}}
\email{guanpeng-li@uiowa.edu}

\author{Shuaiwen Leon Song}{\AFFIL{University of Sydney}{Sydney}{NSW}{Australia}}
\email{shuaiwen.song@sydney.edu.au}

\author{Dingwen Tao}{\AFFIL{Washington State University}{Pullman}{WA}{USA}}
\email{dingwen.tao@wsu.edu}

\begin{abstract}

Deep neural networks (DNNs) are becoming increasingly deeper, wider, and non-linear due to the growing demands on prediction accuracy and analysis quality. Training wide and deep neural networks though requires a large amount of processing time and computing resources such as memory, storage, and I/O. When training a DNN model, the intermediate activation data must be saved in the memory during forward propagation and then restored for backward propagation. However, state-of-the-art accelerators such as GPUs are only equipped with very limited memory capacities due to hardware design constraints, which significantly limits the maximum batch size and hence performance speedup when training large-scale DNNs. Traditional memory saving techniques such as data recomputation and migration either suffers from a high performance overhead or is constrained by specific interconnect technology and limited bandwidth. 

In this paper, we propose a novel memory-driven high performance DNN training framework that leverages error-bounded lossy compression to significantly reduce the memory requirement for training in order to allow training larger neural networks. Different from the state-of-the-art solutions that adopt image-based lossy compressors such as JPEG to compress the activation data, our framework purposely designs error-bounded lossy compression with a strict error-controlling mechanism. Specifically, we provide theoretical analysis on the compression error propagation from the altered activation data to the gradients, and then empirically investigate the impact of altered gradients over the entire training process. Based on these analyses, we then propose an improved lossy compressor and an adaptive scheme to dynamically configure the lossy compression error-bound and adjust the training batch size to further utilize the saved memory space for additional speedup. We evaluate our design against state-of-the-art solutions with four widely-adopted DNNs and the ImageNet dataset. Results demonstrate that our proposed framework can significantly reduce the training memory consumption by up to 13.5$\times$ and 1.8$\times$ over the baseline training and state-of-the-art framework with compression, respectively, with little or no accuracy loss.

\end{abstract}

\maketitle
\pagestyle{plain}

\setlength{\textfloatsep}{6pt}

\section{Introduction}
Deep neural networks (DNNs) have rapidly evolved to the state-of-the-art technique for many artificial intelligence (AI) tasks in various science and technology domains, including image and vision recognition \cite{simonyan2014very}, recommendation systems~\cite{wang2015collaborative}, and natural language processing (NLP)~\cite{collobert2008unified}.
DNNs contain millions of parameters in an unparalleled representation, which is efficient for modeling complexity nonlinearities. 
Many works \cite{krizhevsky2012imagenet,szegedy2015going,he2016deep} have suggested that using either deeper or wider DNNs is an effective way to improve analysis quality, and in fact many recent DNNs have gone significantly deeper and/or wider \cite{wang2018superneurons,jin2019deepsz}. For instance, OpenAI recently published their new DNN-based NLP model GPT-3~\cite{brown2020language} with 175 billion parameters, which is the largest NLP model that is ever trained. Compared with its predecessor GPT-2, GPT-3 expands the capacity by three orders of magnitudes without significant modification to the model architecture, instead just adopting deeper and/or wider layers~\cite{brown2020language}.

DNN training is an optimization process to minimize the target loss function, 
which reflects the event or values into a real number intuitively to represent the cost or difference.
Training modern DNNs requires a large amount of computation and resources such as memory, storage, and I/O. 
A typical strategy to implement efficient DNN training is to rely on advanced high-performance computing (HPC) or data center architectures equipped with high-throughput accelerators, such as general-propose accelerators (e.g., GPUs) or customized hardware (e.g., FPGA or ASIC designs), to perform floating-point operations in a massively parallel manner. 
However, under the current large-scale network models, efficient training is hard to achieve. For example,  training GPT-3 model will take a high-end NVIDIA Tesla V100 GPU 355 years, and cost \$4.6 million on a cost-effective cloud instance~\cite{gtp-price}. This significantly hinders the exploration and adoption of new AI algorithms and DNN designs.   
Furthermore, designing new DNN architectures and training algorithms for various AI tasks require numerous trial-and-error fine-tuning instances, worsening the training cost challenge. 

In this paper, we explore a \textit{general memory-driven approach for enabling efficient DNN training}. Specifically, the ultimate goal is to drastically reduce the memory requirement for training in order to reduce the max batch size limitation for large training speedup. 
When training a DNN model, the intermediate activation data (i.e., the input of all the neurons) is typically saved in the memory during forward propagation, and then restored during backpropagation to calculate gradients and update weights accordingly \cite{hecht1992theory}. However, taking into account the deep and wide layers in the current large-scale nonlinear DNNs, storing these activation data from all the layers requires large memory spaces which are not available in the state-of-the-art training accelerators such as GPUs. For instance, training Inception-V4 \cite{szegedy2017inception} with the batch size of 32 on the ImageNet-2012 dataset \cite{ILSVRC} requires more than 40 GBs of memory, which is larger than the memory available on the state-of-the-art GPUs such as NVIDIA Tesla V100 with 32 GBs. Furthermore, modern DNN model designs trades-off memory requirement for higher accuracy. For example, Gpipe \cite{huang2019gpipe} increases the memory requirement by more than 4$\times$ for achieving a 5\% top-1 accuracy improvement from Inception-V4.

Evolving in recent years, model-parallel \cite{ben2019demystifying} techniques that distribute a model into multiple nodes can reduce the memory consumption of each node but introduce high communication overheads; on the other hand, data-parallel techniques \cite{sergeev2018horovod} remain the same model in every node but distribute the training data to different nodes, thereby suffering from high memory consumption to fully utilize the computational power.
Several techniques such as recomputation, migration, and lossless compression for activation data have been proposed to address the memory consumption challenge for training large-to-large-scale DNNs. For example, GeePS~\cite{cui2016geeps} and vDNN~\cite{rhu2016vdnn} have developed data migration techniques for transferring the intermediate data from GPU to CPU to alleviate the memory burden. However, the performance of data migration approaches is limited by the specific in-node interconnect technology (e.g., PCIe and NVLinks \cite{foley2017ultra}) and its available bandwidth. 
Some other approaches proposed to recompute the activation data  \cite{chen2016training,gomez2017reversible} but they often incur large performance degradation, especially for computationally intensive layers such as convolutional layers. 
Moreover, memory compression approaches based on lossless compression on the activation data \cite{son2014data}) suffers from the limited compression ratio, e.g., only around 2:1 for most floating-point data. 
Alternatively, recent works \cite{evans2020jpeg,choukse2020buddy} proposed to develop compression offloading accelerators for reducing the activation data size before transferring it to the CPU DRAM.  
However, adding a new dedicated hardware component to the existing GPU architecture requires tremendous industry efforts and is not ready for immediate deployment. This design may not be general enough to accommodate future DNN models and accelerator architectures. 

To tackle these challenges, we propose an memory-efficient deep neural network training framework by compressing the activation data using flexible error-bounded lossy compression. 
Compared to lossy compression approaches such as JPEG \cite{wallace1992jpeg} and JPEG2000 \cite{taubman2012jpeg2000}, error-bounded lossy compression can provide a more strict control over the errors occurred to the floating point activation data; compared to lossless compression such as GZIP \cite{gzip} and Zstd \cite{zstd}, it can offer a much higher compression ratio to gain higher memory consumption reduction and performance improvement. 
The key insights explored in this work include:
(i) the impact of compression errors occurred in the activation data on the gradients and the entire DNN training process under the strict error-controlling lossy compression can be theoretically and experimentally analyzed, and
(ii) the training accuracy can be well maintained based on a dynamic fine-grained control over error-bounded lossy compression (i.e., compression error). 
To the best of our knowledge, this is the first work \textit{to theoretically investigate the compression error propagation during DNN training and leverage this analysis to significantly reduce the memory consumption for training large DNNs while maintaining the high training accuracy}.
In summary, this paper makes the following contributions:

\begin{itemize}
    \item We propose a novel memory-efficient DNN training framework via dynamically compressing the intermediate activation data through fine-grained error-bounded lossy compression.
    \item We provide a thorough analysis on the impact of compression error propagation during DNN training from both theoretical and empirical perspectives. Moreover, we improve the lossy compression algorithm to avoid the significant alteration (vanish or explosion) of gradients due to the errors introduced to the activation data. 
    \item We propose an adaptive scheme to dynamically configure the error-bounded lossy compression based on a series of current training status data.
    \item We evaluate our proposed training framework on four widely-adopted DNN models (AlexNet, VGG-16, ResNet-18, ResNet-50) with the ImageNet-2012 dataset and compare it against state-of-the-art solutions. Experimental results show that our design can reduce the memory consumption by up to 13.5$\times$ and 1.8$\times$ compared to the original training framework and the state-of-the-art method, respectively, under the same batch size. 
\end{itemize}

\section{Background and Motivation}
\label{sec:background}

In this section, we first present the background information on training large-scale DNNs and error-bounded lossy compression for floating-point data. We then discuss the motivation of this work and research challenges.

\subsection{Training Large-Scale DNNs}

DNNs have been widely studied in recent years, which have proven their capabilities in various science domains such as vision detection \cite{russakovsky2015imagenet} and natural language processing (NLP) \cite{young2018recent}. Each neural network contains many connected layers that allow input data flow to compute through and provide the associated results. There are different types of layers that have been developed for DNNs, among which convolutional layer and its optimization have become one of the research focuses for the deep learning (DL) community in the recent years, e.g., the invention of TPU \cite{googletpu}. They consist of a set of filters and kernels that perform convolutions with a sliding window on the input data.
Convolutional neural networks (CNNs) typically consist of convolutional, activation, pooling, batch normalization, local response normalization, fully connected, and dropout layers. 
To further improve the state-of-the-art accuracy on large real-world datasets, recent works \cite{he2016deep, szegedy2017inception} have proposed deeper and wider nonlinear CNN architectures, e.g., ResNet \cite{he2016deep} and Inception V4 \cite{szegedy2017inception}. 
For instance, ResNet-50 with 50 convolutional layers can provides 71.49\% top-1 accuracy for the ImageNet-12 dataset, but the early linear CNN architecture like AlexNet \cite{krizhevsky2012imagenet} with just five sequential layers can only provide 57.41\% top-1 accuracy.

\begin{figure}[]
    \centering
    \includegraphics[width=1.0\linewidth]{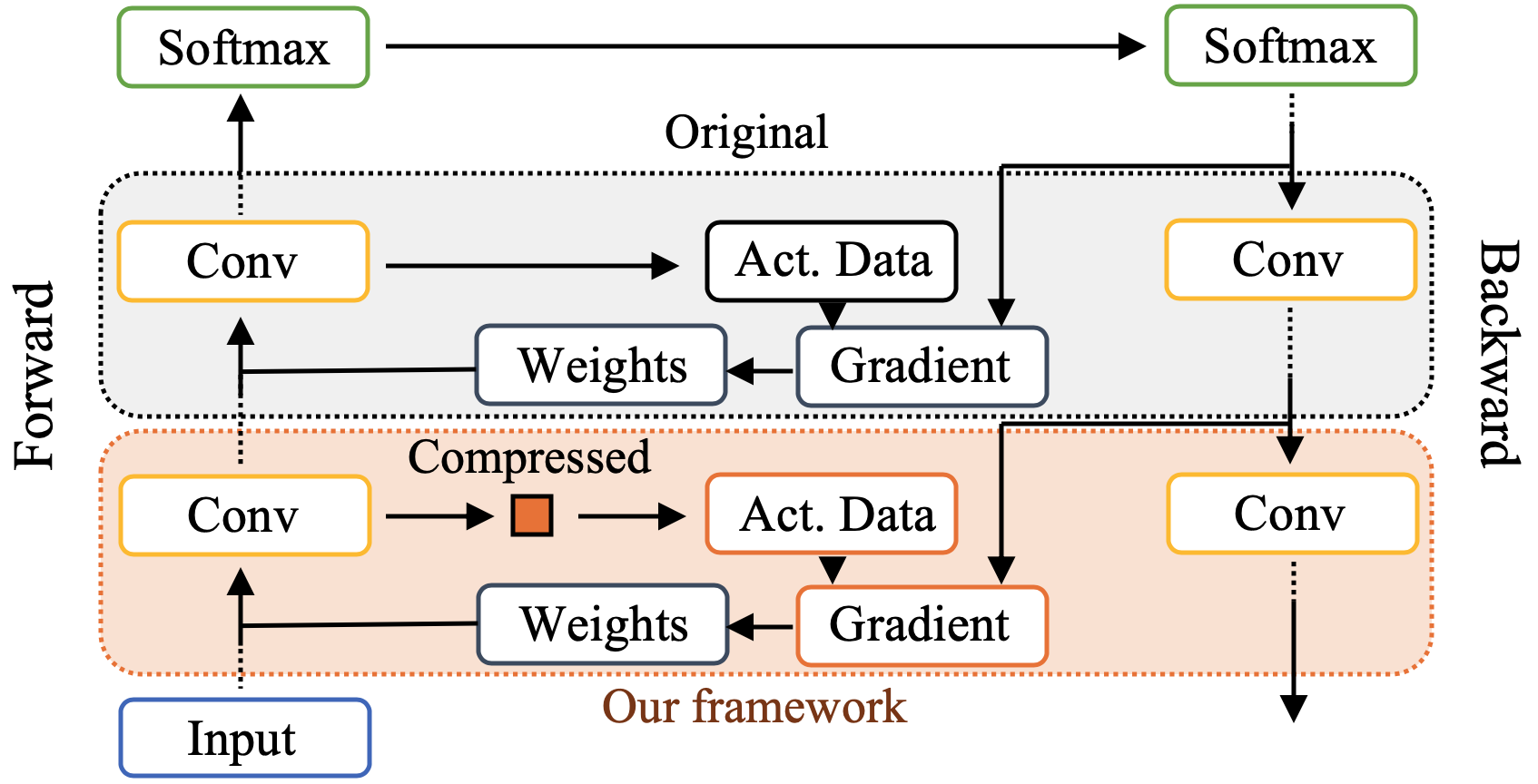}
    \caption{Data flow in a sample iteration of training CNNs.}
    \label{fig:fig-1}
\end{figure}

\begin{figure}[]
    \centering
    \includegraphics[width=0.87\linewidth]{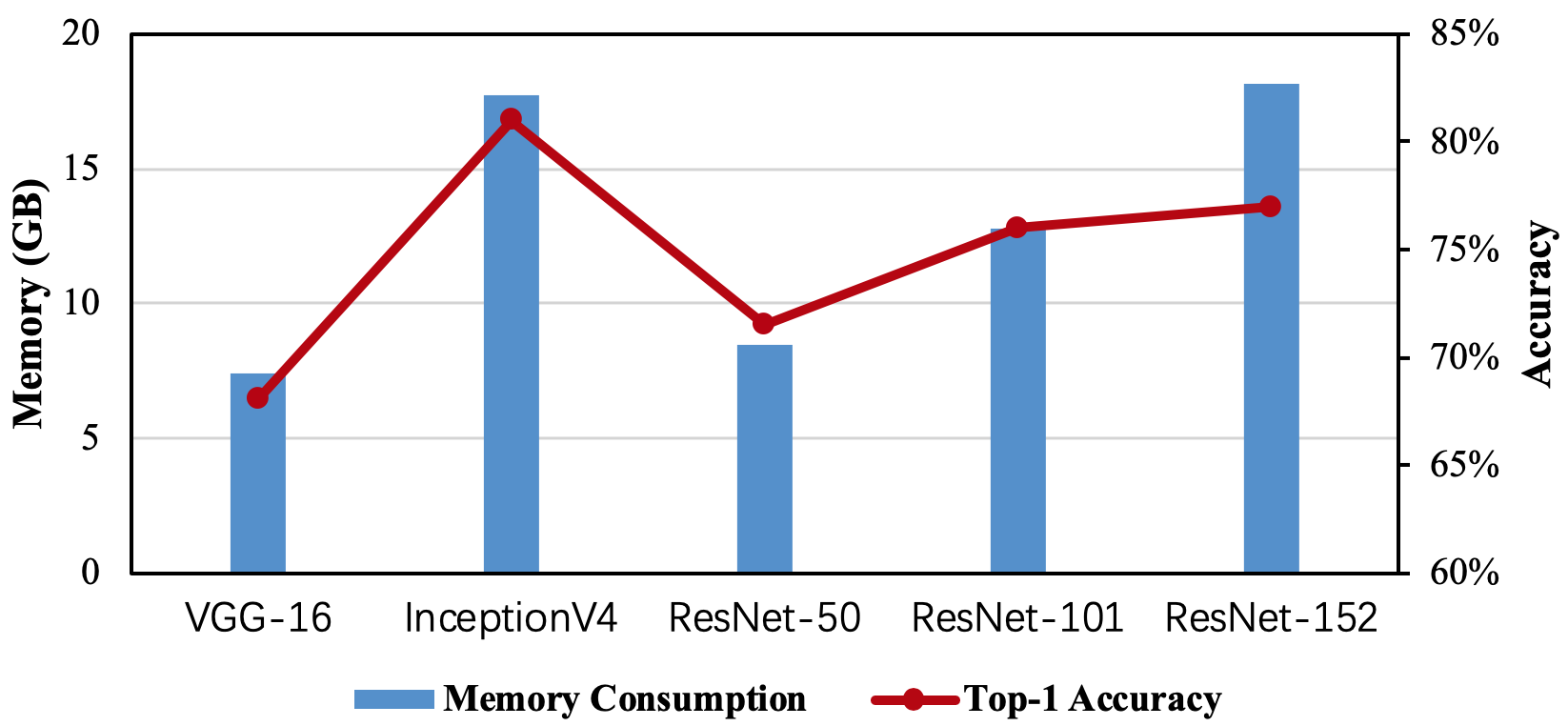}
    \caption{Memory consumption and top-1 accuracy of different state-of-the-art neural networks.}
    \label{fig:fig-2}
\end{figure}

Training deep and wide neural networks has become increasingly challenging. While many state-of-the-art deep learning frameworks such as TensorFlow \cite{abadi2016tensorflow} and PyTorch \cite{paszke2019pytorch} can provide high computation throughput by leveraging the massive parallelism on general-purpose accelerators such as GPUs, one of the most common bottlenecks remains to be the high memory consumption during the training process, especially considering the limited on-chip memory available on modern DNN accelerators. This is mainly due to the ever-increasing size of the activation data computed in the training process. Training a neural network involves many training iterations/epochs to fine tune the model weights. Each iteration includes a forward and backward propagation, as shown in Figure~\ref{fig:fig-1}. The intermediate activation data (the output from each neuron) generated by every layer are commonly kept in the memory until the backpropagation reaches this layer again. Several works \cite{chen2016training, gomez2017reversible,rhu2016vdnn, cui2016geeps,evans2020jpeg} have pointed out that there would be a huge gap between the time when the activation data is generated in the forward propagation and the time when the activation data is reused in the back propagation, especially when training very deep and wide neural networks.
Figure~\ref{fig:fig-2} shows the memory consumption of various neural networks. For these CNNs, comparing with the model/weight size, activation data size is much larger, since the convolution kernels are relatively small compared to the activation tensors. 
In summary, we are facing two main challenges due to the high memory consumption in today's deep learning training: 
(1) it is challenging to scale up the training process under a limited GPU memory capacity; and 
(2) a limited batch size leads to low training performance.
The performance challenge is further exacerbated when the error-and-trail based approach is typically used in practice for developing new DNN models and algorithms.

In recent years, several works have been proposed to reduce the memory consumption for DNN training, including activation data recomputation\cite{chen2016training, gomez2017reversible}, migration \cite{rhu2016vdnn, cui2016geeps}, and compression \cite{evans2020jpeg}. 
Recomputation takes advantage of the layers with low computational cost, such as pooling layer. Specifically, it would deallocate the activation data for those layers and recompute them based on their prior layer during the back propagation when needed. 
This method can reduce some unnecessary memory cost, but it can only be applied to limited types of layers with low performance overhead.
Layers such as convolutional layers often cannot be recomputed efficiently with acceptable performance overhead. 

Another type of methods are proposed around data migration \cite{rhu2016vdnn, cui2016geeps}, which sends the activation data from the accelerator to the CPU host when generated, and then loads it back from the host when needed. 
However, the performance of data migration heavily depends on the interconnect bandwidth available between the host and the accelerator(s), and the in-node interconnect technology applied. For example, NVLink \cite{foley2017ultra} technology is currently limited to high-end NVIDIA AI nodes (e.g., DGX series) and IBM power series. 
This paper targets to develop a general technique that can be applied to all types of HPC and datacenter systems. 

Last but not least, data compression is another efficient approach to reduce the memory consumption, especially for conserving the memory bandwidth \cite{evans2020jpeg, lal20172mc, ekman2005robust}. The basic idea using data compression here is to compress the activation data when generated, hold the compressed data in the memory, and decompress it when needed. 
However, using lossless compression \cite{choukse2020buddy} can only provide a relatively low memory reduction ratio (i.e., compression ratio), e.g., typically within 2$\times$. 
Some other studies such as JPEG-ACT \cite{evans2020jpeg} leverages the similarity between activation tensors and images for vision recognition tasks and apply a modified JPEG lossy compressor to activation data. This approach provides a higher compression ratio (e.g., typically around 7$\times$) compared to the lossless compression based solution. But it suffers from two main drawbacks.
First, it introduces uncontrollable compression error to activation data and eventually could lose the control of the overall training accuracy, since JPEG is mainly designed for images and is an integer based lossy compression. The JPEG based solution (or other image based lossy compression) may not be suitable for DNNs applied in a more general and larger scientific context than image processing to analyze extremely large amounts of data, e.g., large-scale HPC scenarios for unprecedented discoveries \cite{wozniak2018candle} via deep learning. 
Second, the JPEG based solution \cite{evans2020jpeg} needs a support from a dedicated hardware component to be added in GPU hardware. Hardware innovation typically takes 5 to 10 years for adaptation and it cannot be directly deployed to today's systems.

We note that all the three methods above are orthogonal to each other, which means they could be deployed together to maximize the compression ratio and training performance. Thus, in this paper, we mainly focus on designing an efficient data compression, more specifically lossy compression based solution, to achieve the memory reduction ratio beyond the state-of-the-art approach on CNN models. 
In addition, since convolutional layers are the most difficult type of layers for efficient recomputation, our solution focuses on the convolutional layers to provide high compression ratio with minimum performance overhead and accuracy loss.  

\subsection{Error-Bounded Lossy Compression}
Floating-point data compression has been studied for decades. There are two main categories: lossless compression and lossy compression. Lossless compressors such as FPZIP~\cite{lindstrom2006fast} and FPC~\cite{FPC} can only provide limited compression ratios (typically up to 2:1 for most scientific data) due to the significant randomness of the ending mantissa bits~\cite{son2014data}.

Lossy compression, on the other hand, can compress data with little information loss in the reconstructed data.
Compared to lossless compression, lossy compression can provide a much higher compression ratio while still maintaining useful information for scientific discoveries. 
Different lossy compressors can provide different compression modes, such as error-bounded mode and fixed-rate mode. 
Error-bounded mode requires users to set an error bound, such as absolute error bound and point-wise relative error bound. The compressor ensures the differences between the original data and the reconstructed data do not exceed the user-set error bound.
Fixed-rate mode means that users can set a target bitrate, 
and the compressor guarantees the actual bitrate of the compressed data to be lower than the user-set value.

In recent years, a new generation of lossy compressors for scientific data have been proposed and developed, such as SZ~\cite{di2016fast, tao2017significantly, liangerror} and ZFP~\cite{zfp}.
Unlike traditional lossy compressors such as JPEG \cite{wallace1992jpeg} which are designed for images (in integers), SZ and ZFP are designed to compress floating-point data and can provide a strict error-controlling scheme based on user's requirements.
In this work, we chose to use SZ instead of ZFP because the GPU version of SZ---cuSZ \cite{tian2020cusz}\footnote{Compared to CPU SZ, cuSZ can provide much higher compression and decompression throughput on GPUs.}---provides a higher compression ratio than ZFP and offers the absolute error bound mode that ZFP does not support (but necessary for our error control).
Specifically, SZ is a prediction based error-bounded lossy compressor for scientific data. SZ has three main steps: (1) predict each data point's value based on its neighboring points by using an adaptive, best-fit prediction method; (2) quantize the difference between the real value and predicted value based on the user-set error bound; and (3) apply a customized Huffman coding and lossless compression to achieve a higher compression ratio.

\subsection{Research Goals and Challenges}

In this paper, we propose a novel memory-driven high performance CNN training framework that leverages error-bounded lossy compression to significantly reduce the memory requirement for training in order to allow larger neural network training and developing. 
To achieve this goal, there are several critical challenges to be addressed. 
First, since we plan to use an error-bounded lossy compressor, a strictly controlled compression error would be introduced to the activation data.
Second, in order to maintain the training accuracy curve with a minimum impact to the performance and final model accuracy, \emph{we must understand how the introduced error would propagate through the whole training process.} In other words, we must theoretically and/or experimentally analyze the error propagation, which is challenging. To the best of our knowledge, there is no prior investigation on this.

Third, once we understand the connection between the controlled error and training accuracy, how to balance the compression ratio and accuracy degradation in a fine granularity is also challenging. 
In other words, a more aggressive compression can provide a higher compression ratio but also introduces more errors to activation data, which may significantly degrades the final model accuracy or training performance (cant converge). 
Thus, we must find a balance to offer as high compression ratio as possible to different layers across different iterations while maintaining minimum impact to the accuracy. 
\section{Compression Error Propagation Analysis}
\label{sec:analysis}

In this section, we present the theoretical support of our proposed training framework---analyzing compression error propagation (1) from activation data to gradient and (2) from gradient to overall training for convolutional layers. 

\subsection{Modeling Compression Error}
\label{subsec:model}

cuSZ \cite{tian2020cusz} is a prediction based error-bounded lossy compressor for floating-point data on GPUs.
It first uses a dual-quantization technique to quantize the floating-point input data based on user-set error bound. Then, it applies Lorenzo based predictor to efficiently predict the value of each data point based on its neighboring points.  After that, a quantization code (integer) is generated for each value. Finally, a customized Huffman coding is applied to all the quantization codes. Similar to the original SZ, the error introduced to the input data after decompression usually forms an uniform distribution. This is mainly because of the linear-scaling quantization technique adopted. We refer readers to \cite{lindstrom2017error} for more details about the error distribution of SZ from a statistical perspective. 

\begin{figure}[]
    \centering
    \includegraphics[width=0.7\linewidth]{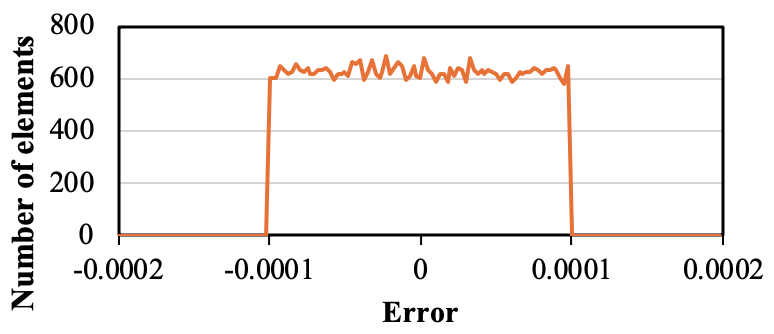}
    \vspace{-2mm}
    \caption{Sample error distribution of activation data compressed by cuSZ lossy compression with error bound $10^{-4}$.}
    \label{fig:fig-31-1}
\end{figure}

Figure~\ref{fig:fig-31-1} illustrates an example error distribution when compressing and decompressing an activation data by cuSZ. 
The activation data used here for demonstration is extracted from Conv-5 layer of AlexNet \cite{krizhevsky2012imagenet} in a certain iteration. 
Note that in fact we plot the error distribution every 50 iterations, and observe that all the error distributions are quite similar and follow uniform distribution, which is consistent with the conclusion drawn in the prior work \cite{lindstrom2017error}. 
Thus, we propose to use the uniformly distributed error model for the theoretical analysis and an error injection based approach to demonstrate the effectiveness of our theoretical derivation in this section. 
Note that for the purpose of theoretical analysis, we inject the error, rather than actually compressing and decompressing activation data, to demonstrate how uniformly distributed error propagates from activation data to gradient and then to the whole training process. We leave the evaluation of actual compression to Section \ref{sec:evaluation}. 

\subsection{Modeling Error Impact on Gradient}
\label{sub:theory-analysis}

Next, we theoretically derive how error propagates from activation data to gradient and provide an experimental proof based on statistical analysis using error injection.

\begin{figure}[]
    \centering
    \includegraphics[width=0.95\linewidth]{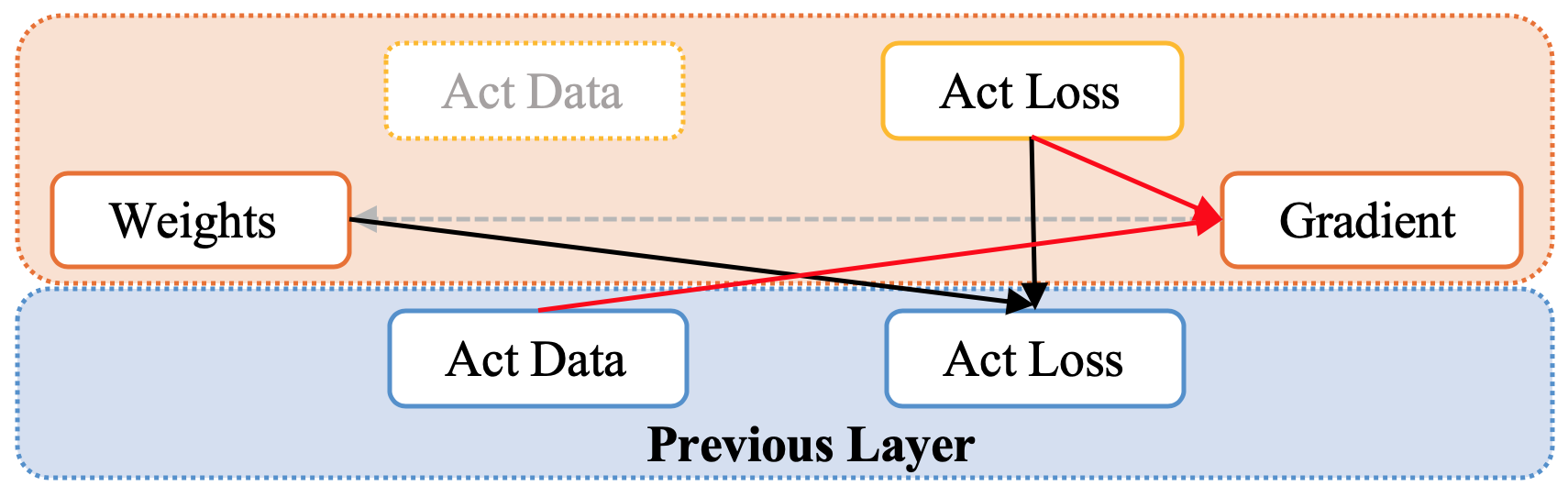}
    \vspace{-2mm}
    \caption{Data dependencies in one convolutional layer during backpropagation.}
    \label{fig:fig-32-3}
    \vspace{-2mm}
\end{figure}

As shown in Figure~\ref{fig:fig-1}, the compressed activation data needs to be decompressed when the back propagation reaches the corresponding layer. During the back propagation, each layer computes the gradient to update the weights and the loss to be propagated to the previous layer (back propagation). 
As shown in Figure~\ref{fig:fig-32-3}, on the one hand, the loss of activation data for the previous layer only depends on the current layer's loss and weight. 
On the other hand, the gradient depends not only on the loss of the current layer, but also on the activation data.
We note that the activation data can be compressed during the forward propagation to reduce the memory consumption. 
In a conclusion, in order to understand the impact of compressing the activation data, we must first understand how compression error introduced to the activation data would propagate to the gradient.

\begin{figure}[]
	\begin{subfigure}{\linewidth}\centering
	    \includegraphics[width=0.87\linewidth]{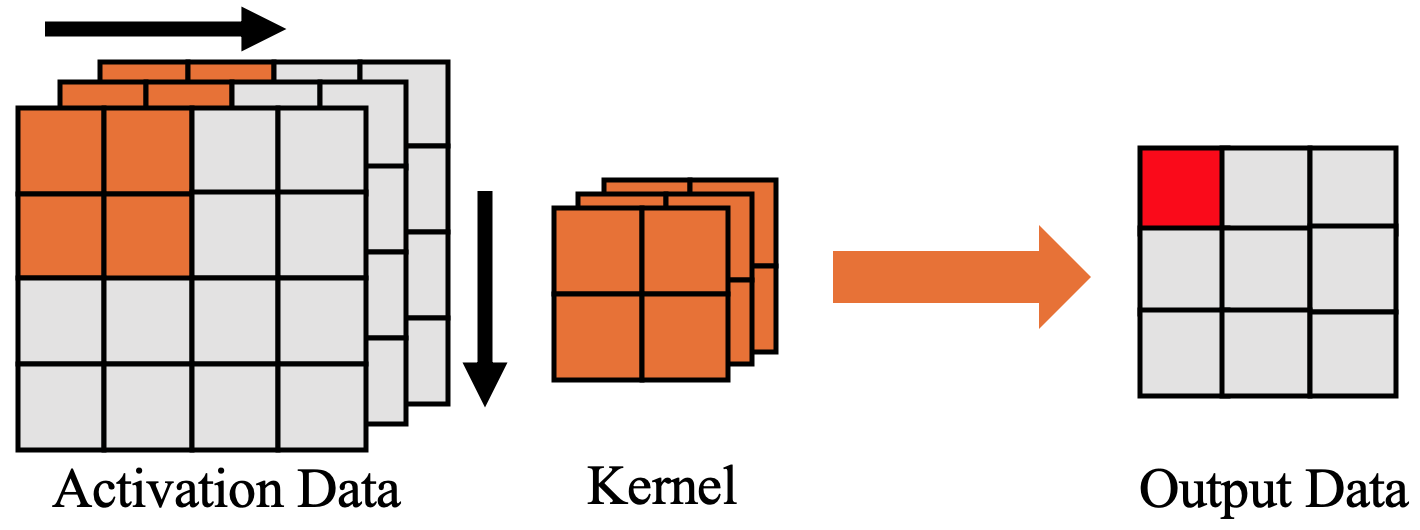}
	    \caption{\footnotesize Forward pass}\label{fig:subfigure_dependency_forward}
	\end{subfigure}
	\begin{subfigure}{\linewidth}\centering
	    \includegraphics[width=0.87\linewidth]{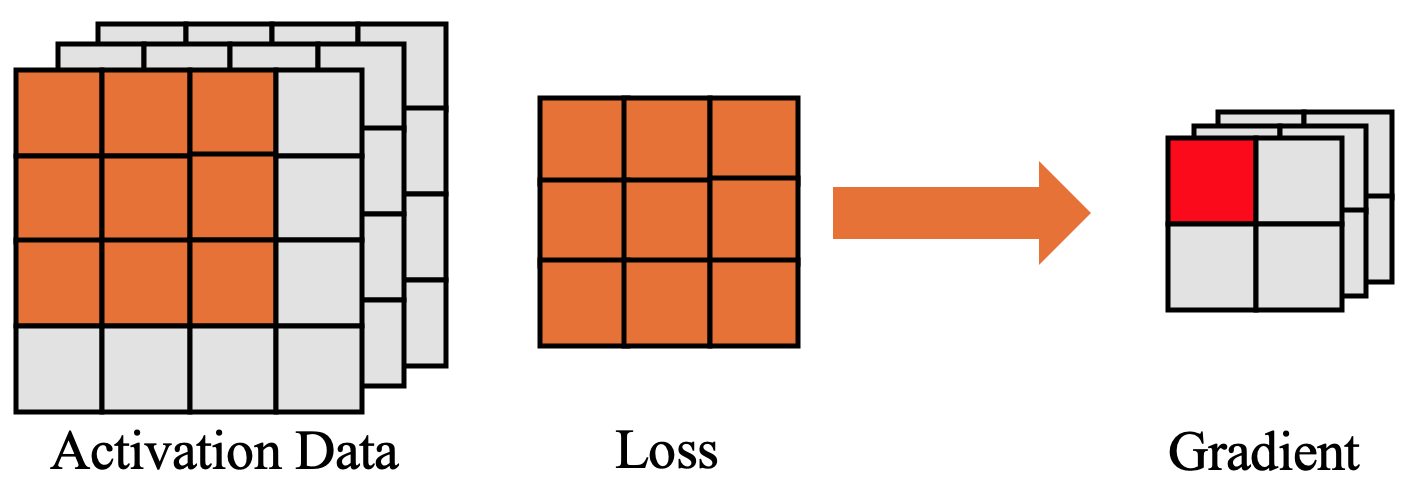}
	    \caption{\footnotesize Backward pass}\label{fig:subfigure_dependency_backward}
	\end{subfigure}
\vspace{-4mm}
\caption{Forward and backward computation in a sample convolutional layer. Kernel size is 2$\times$2, stride is 1, input channel count is 3, and output channel count is 1.}
\label{fig:fig-32-4}
\vspace{-2mm}
\end{figure}

In the forward pass, multiple kernels perform convolutions on the input activation data. 
As shown in Figure~\ref{fig:subfigure_dependency_forward}, the kernel is performed on the activation data (marked in brown) and generates the output value as shown in red. 
Similar to forward pass, the backward pass reverses the computation, where the parameter's gradient is computed based on the loss (with the same dimension of the output data in the forward pass) and the original data in the kernel, as shown in Figure~\ref{fig:subfigure_dependency_backward}.
Similarly, the activation data and the loss compute the gradient value as shown in red. More specifically, this value is computed as
\begin{equation}
    G_{k,i} = A_{k,0} {\times} L_{0} + A_{k,1} {\times} L_{1} + ... + A_{k,n^{'}} {\times} L_{n},
    \label{equ-1}
\end{equation}
where $G$ is the gradient, $A$ is the activation data, $L$ is the loss, $k$ is the current channel, $i$ is the value index of the channel, $n$ is the number of values in the loss matrix, $n^{'}$ is the corresponding index of activation data to loss matrix. Note that for simplicity, we ignore all the $i$s on the right side of Equation (\ref{equ-1}). 
Note that here, if the number of output channels is greater than 1, which is true for most convolutional layers, the same process will be used for multiple kernels, as shown in Figure~\ref{fig:subfigure_dependency_backward}, and the above formula still holds in this case.

Based on our analysis in Section \ref{subsec:model}, the error introduced to the activation data is uniformly distributed, 
thus, in the backpropagation, we can have
\begin{align}
    G^{'}_{k,i} &= A^{'}_{k,0} {\times} L_{0} + A^{'}_{k,1} {\times} L_{1} + ... + A^{'}_{k,n^{'}} {\times} L_{n}, \label{equ-2} \\
    A^{'}_{k,j} &= A_{k,j} + e, e \in [-eb, +eb], \notag
\end{align}
where $A^{'}$ is the decompressed activation data,
$G^{'}$ is the gradient altered by the compression error, $e$ is the error, and $eb$ is the user-set absolute error bound. After a simple transformation, we can have
\begin{align}
    G^{'}_{k,i} &= A_{k,0} {\times} L_{0} + A_{k,1} {\times} L_{1} + ... + A_{k,n^{'}} {\times} L_{n}\notag \\ &\quad + e_0 {\times} L_{0} + e_1 {\times} L_{0} + ... + e_n {\times} L_{n}\notag \\
    &= G_{k,i} + E, \label{equ-3} \\
    E &= e_0 {\times} L_{0} + e_1 {\times} L_{1} + ... + e_n {\times} L_{n},\notag
\end{align}
where $E$ is the gradient error. 

Although it is not possible to calculate or predict the exact value of every element $E$, we can predict its distribution based on our previous assumption. 
We also note that the batch size of a typical neural network is usually relatively large during the training process, such as 128$\sim$256, since larger batch size results in higher training performance in general.
As a result, the final gradient for updating weights can be computed as follows. 
\begin{align}
    G^{'}_{final} &= Average(G^{'}_0, G^{'}_1, ... , G^{'}_N), \notag\\
    E_{final} &= Average(E_0, E_1, ... , E_N) \label{equ-4} \\
    &= ( e_{0,0} {\times} L_{0,0} + e_{0,1} {\times} L_{0,1} + ... + e_{0,n} {\times} L_{0,n} \notag \\
    & \quad + e_{1,0} {\times} L_{1,0} + e_{1,1} {\times} L_{1,1} + ... + e_{1,n} {\times} L_{1,n}\notag \\
    & \quad \dots \notag \\
    & \quad + e_{N,0} {\times} L_{N,0} + e_{N,1} {\times} L_{N,1} + ... + e_{N,n} {\times} L_{N,n} )/N, \notag \\
    e &\in [-eb, +eb], \notag
\end{align}
where $N$ is the number of batch size. 
Note that all $e$ are independently and uniformly distributed as discussed in Section \ref{subsec:model}. Although $L$ can be related with each other in the same batch, they are still independently distributed across different batches. 
According to \cite{wiki}, the sum of a series independent random variables with the same distribution follows a normal distribution,
which means the error distribution of the gradient can be expected to be be normally distributed. 
We identify that the distribution of loss $L$ for one input (i.e., one image) is highly concentrated in zero, where the highest value in the loss is usually much larger than the average of $L$. Thus, we can simplify our Equation \ref{equ-4} to 
\begin{align}
    E_{final} & \approx e_{0} {\times} L_{max, 0} + e_{1} {\times} L_{max, 1} + ... + e_{N} {\times} L_{max, N}, \label{equ-8} \\
    e &\in [-eb, +eb], \notag
\end{align}
where $L_{max}$ is the maximum value in the loss for each input alone. In order to reduce the complexity, we can greatly improve the performance with fewer parameters that need to be collected.

\begin{figure}[]
    \centering
    \begin{subfigure}{\linewidth}
    \centering
    \includegraphics[width=0.95\linewidth]{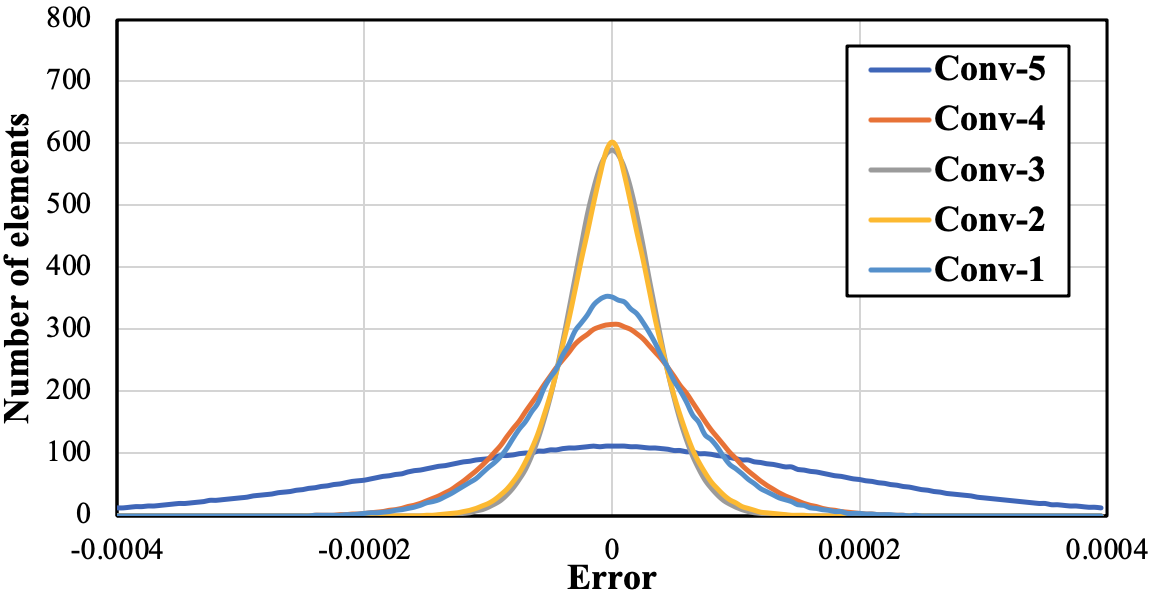}
    \caption{\footnotesize Original zeros been compressed}
    \label{fig:fig-32-1}
    \end{subfigure}

    \begin{subfigure}{\linewidth}
    \centering
    \includegraphics[width=0.95\linewidth]{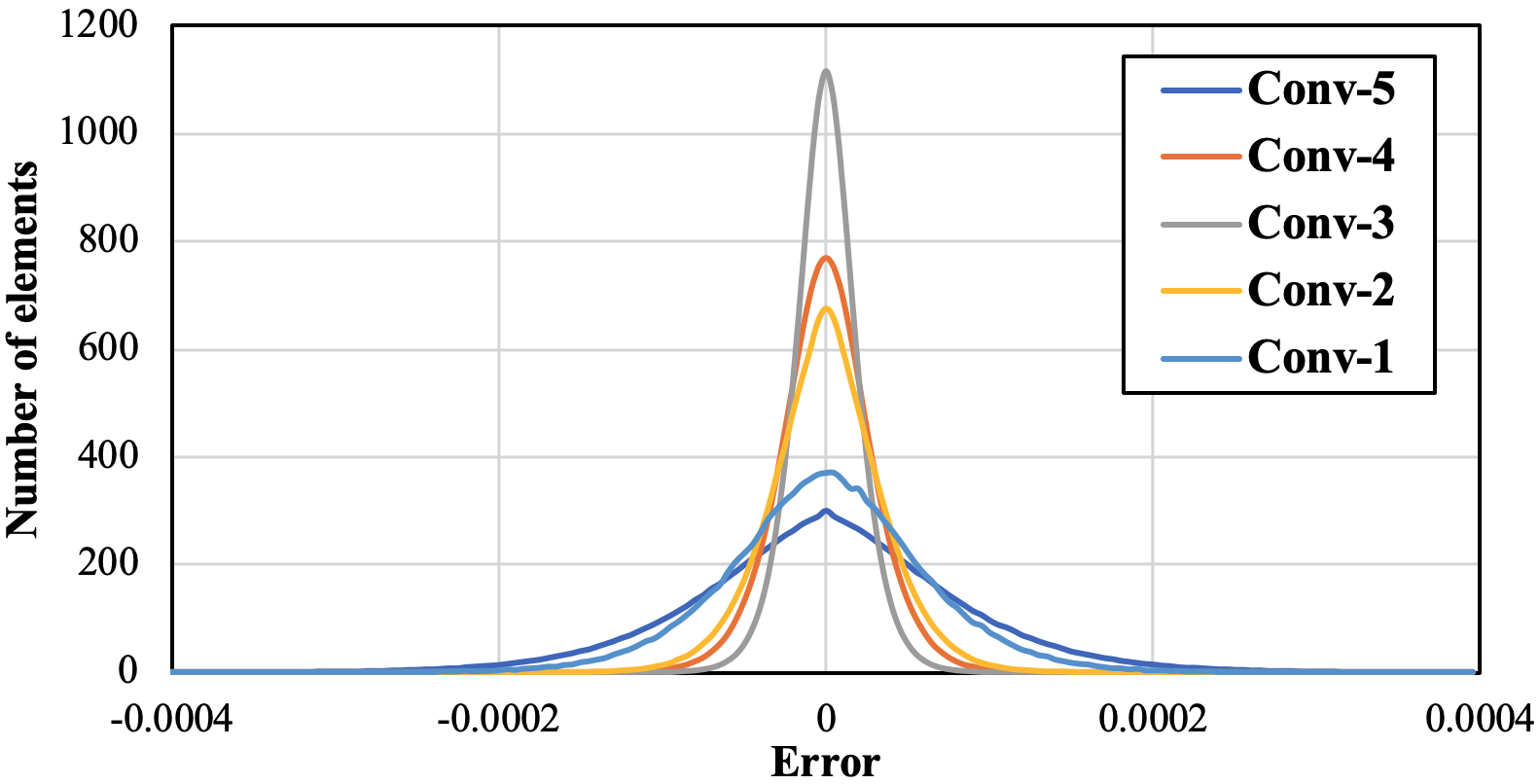}
    \caption{\footnotesize Original zeros remain zero}
    \label{fig:fig-32-2}
    \end{subfigure}
    \vspace{-2mm}
    \caption{Distributions of gradient error when injecting modeled compression error to activation data. Note that number of elements is normalized to ensure different layers under the same scale. Data is collected every 100 iterations.}
    \label{fig:fig-32}
\end{figure}

Next, we inject error (to simulate the compression error) to the activation data in convolutional layers based on our error model discussed in Section \ref{subsec:model}. Then, we collect the error of the gradient in the backpropagation. 

Figure~\ref{fig:fig-32-1} illustrates the normalized error distribution of gradients collected from different layers, all of which follow the normal distribution as expected. 
In fact, by calculating the percentage of the area within $\pm\sigma$ of each curve, we can get a value close to 68.2\%, which confirms our theoretical derivation. 
After that, we need to figure out how to predict $\sigma$ before compression in order to calculate the desired and acceptable error bound for each layer's activation data.

First, we note that $\sigma$ is highly related to the number of elements that are combined together. In general, more elements result in larger $\sigma$, vice versa.
A $2\times$ increase of elements results in $\sqrt{2}\times$ increase of $\sigma$, which means that more uncertainties has been added to the system. 
Second, $\sigma$ is also related to the value scale, in this case, the average of the loss $L$ at the current layer. 
Note that it is not necessary to compute the average value of the loss every iteration. Instead, we can compute it every $W$ iterations to reduce the computation overhead, since this value is relatively stable in a fixed period of time. 
Based on these two parameters, we can estimate $\sigma$ by the following equation:
\begin{equation}
    \sigma \approx a \bar L \sqrt{N} eb,
    \label{equ-5}
\end{equation}
where $L$ is the loss matrix, $N$ is the batch size, and $a$ is an empirical coefficient. Note that this coefficient $a$ is unchanged for different neural networks because it is essentially a simplified value of the previous equation.

Finally, we note that there is a noticeable fraction of activation data to be zeros, however, our above analysis so far does not cover it. 
When compressing a series of continuous zeros, cuSZ may change them into a small value within the user-set error bound instead of following the error distribution. 
In such case, we manually set these values (lower than the error bound) to zeros when decompression,
meaning no error would be introduced when facing continuous zeros. 
Moreover, in some cases, the activation data may contain many zeros due to the activation function layer (such as the ReLU layer) before the current convolutional layer. If this happens, the activation data of convolutional layer can be quickly recomputed through the activation function instead of being saved, which will essentially erase the negative values to zeros. 
Since lossy compression such as cuSZ is unlikely to change the sign of the activation data value, these data will remain at zero.

Figure~\ref{fig:fig-32-2} shows the distribution of the gradient error after we inject the error to the activation data (maintaining zeros unchanged). Compared with Figure~\ref{fig:fig-32-1}, we can observe a decrease of $\sigma$, but it still holds a normal distribution. This decrease is partly because of the reduction of the number of elements in Equation~\ref{equ-4}, since those zeros do not have any error.
In these cases, we can revise the prediction of $\sigma$ accordingly by the following equation:

\begin{equation}
    \sigma^{'} = \sigma \sqrt{R},
    \label{equ-9}
\end{equation}
where $R$ is the ratio of non-zero elements percentage in the activation data. Again, in practice, we do not need to compute this ratio every iteration but every $W$ iterations, since this ratio is relatively stable in a fixed period of time.

\subsection{Error Propagation Analysis}

\label{sub:exp-analysis}

Finally, we discuss the error propagated from gradient to overall training curve using an experimental analysis. 

Our goal is to identify the maximum acceptable gradient error that would cause little or no accuracy loss.
According to our theoretical and experimental proofs in Section \ref{sub:theory-analysis}, the gradient error can be modeled as normally distributed error. 
In this section, we follow the same strategy used in the last section to inject error to the gradient that follows our error model and perform the analysis and evaluation. 

We propose to use momentum for alleviating the impact of the gradient error. 
Momentum has been widely used in most of neural network training \cite{dong2018boosting}. 
Actually, in order to update the weights, it is based not only on the gradient computed from the current iteration but also on the momentum.
In other words, both the gradient and the momentum (with the same dimension as the weights and gradient) take up a portion of the updated data for weights. 
Thus, it is critical to maintain an accurate momentum vector similar to the error-free one to guide the direction of weight update. 

\begin{figure*}[]
    \centering
    \includegraphics[width=0.91\linewidth]{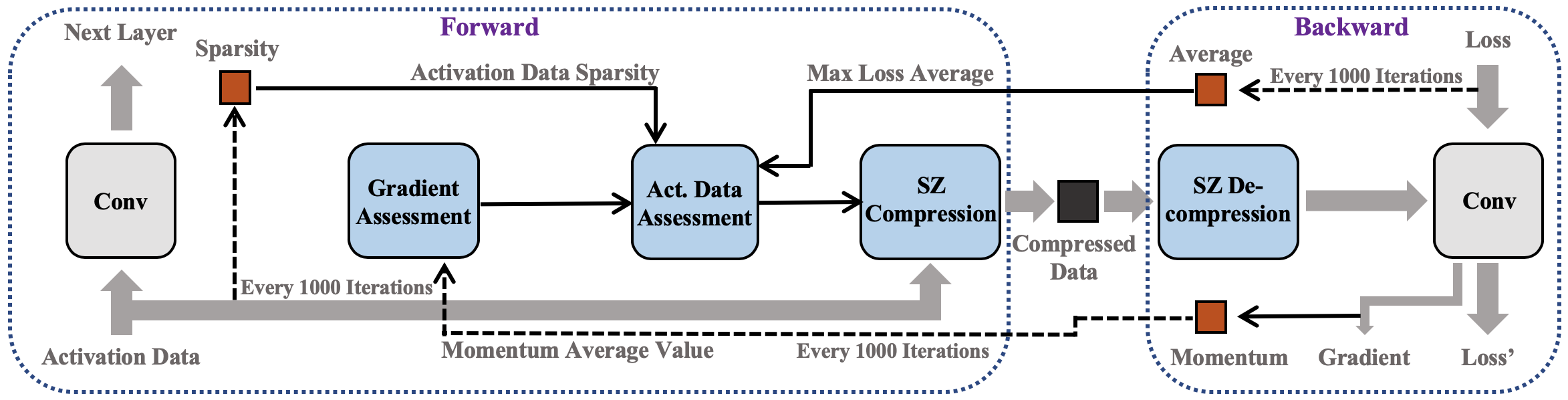}
    \vspace{-4mm}
    \caption{Overview of our proposed memory-efficient CNN training framework.}
    \label{fig:fig-41-1}
\end{figure*}

In fact, the momentum is computed based on the current layer's gradient, and the momentum error is directly correlated to the gradient error. 
While thanks to the normally distributed gradient error, which is centralized and symmetric in the original direction, the momentum error is relatively low compared to the gradient error. 
Therefore, this does help the training towards the correct direction---even a few iterations may generate the undesired gradient, they can be offset quickly through the momentum based on the estimated gradient error.

\section{Design Methodology}
\label{sec:design}

In this section, we describe the overall design of our proposed lossy compression supported CNN training framework and analyze the performance overhead.

Our proposed memory driven framework is shown in Figure~\ref{fig:fig-41-1} and Figure~\ref{fig:fig-1}. We iteratively repeat the process shown in Figure~\ref{fig:fig-41-1} for each convolutional layer in every iteration. Our proposed framework mainly include four phases: (1) parameter collection of current training status for adaptive compression, (2) gradient assessment to determine the maximum acceptable gradient error, (3) estimation of compression configuration (e.g, absolute error bound), and (4) compression/decompression of activation data with cuSZ.

\subsection{Parameter Collection}

First, we collect the parameters of current training status for the following adjustment of lossy compression configurations. 
Our framework mainly collects two types of parameters: (1) offline parameters in CNN architecture, and (2) semi-online parameters including activation data samples, gradient, and momentum.

First of all, we collect multiple static parameters including batch size, activation data size of each convolutional layer, and the size of its output layer. We need these parameters because they affect the number of elements combined into each value in the gradient and hence affect $\sigma$ in its normal error distribution, as discussed in Section \ref{sec:analysis}. It would also help the framework collects corresponding online parameters.

For the semi-online parameters, we collect the sparsity of activation data and the its average loss in backpropagation to estimate how the compression error would propagate from the activation data to the gradient.
For the gradient, we compute the average value of its momentum. 
Note that in many DNN training framework such as Caffe \cite{jia2014caffe} and TensorFlow \cite{abadi2016tensorflow}, momentum is naturally supported and activated, so it can be easily accessed. The data collection phase is shown as the dark thin arrows in Figure~\ref{fig:fig-41-1}.
Note that we still use decompressed data in training for collecting these parameters here.

Moreover, an active factor $W$ needs to be set in the beginning of training process to adjust the overall activeness in our framework.
$W$ is used to determine the activeness of our parameter extraction. 
We only extract semi-online parameters every $W$ iterations to reduce the computation overhead and improve the overall training performance. 
As discussed in Section \ref{sec:analysis}, these parameters vary relatively slowly during the training, thus, we only need to estimate the error impact in a fixed iteration interval in our framework. 
In this paper, we set $W$ to 1000 as default, which provides high accuracy and low overhead in our evaluation.

\subsection{Gradient Assessment}

Next, we estimate the limit of the gradient error that would result in little or no accuracy loss to the training curve,
as shown in Figure~\ref{fig:fig-41-1}. 
Even with the help of the offset from momentum, we still want to keep the gradient of each iteration as close as possible to the original one. Based on our previous analysis in Section \ref{sec:analysis}, we need to determine the acceptable $\sigma$ in the gradient error distribution that minimizes the impact to the overall training accuracy curve. 
We use 1\% as the acceptable error rate based on our empirical study, i.e., the $\sigma$ in the momentum error model needs to be 
\begin{align}
    \sigma = 0.01 M_{Average}
    \label{equ-7} 
\end{align}
where $M_{Average}$ is the average value of the momentum. Note that here we use the average instead of the modulus length of the momentum because we focus on each individual value of the gradient, and the average value can be more representative. Also, note that we do not want to estimate the average value of gradient and momentum; 
instead, we seek an empirical solution to directly monitor the average values. Based on our experiment, the gradient does not tend to vary dramatically in a short time period during training.

\subsection{Activation Assessment}

After that, we dynamically configure the lossy compression for activation data based on the gradient assessment in the previous phase and the collected parameters as shown in Figure~\ref{fig:fig-41-1}. As discussed in Section \ref{sec:analysis}, we need $\sigma$ (from gradient error model), $R$ (sparsity of activation data), $\bar{L}$ (average value of current loss), and $N$ (batch size) to determine the acceptable error bound for compressing the activation data at current layer. We simplify our estimator to the following:
\begin{equation}
    eb = \frac{\sigma}{a \bar{L} \sqrt{N R}}
    \label{equ-8}
\end{equation}
where $eb$ is the absolute error bound for activation data, $\sigma$ describes the acceptable error distribution in the gradient, $a$ is the empirical coefficient, $\bar{L}$ is the average value of  current layer's loss, $N$ is the batch size, and $R$ is the sparsity ratio of activation data.

\subsection{Adaptive Compression}
\label{sub:comp}

In the last phase, we deploy the lossy compression with our optimized configuration to the corresponding convolutional layers. We also monitor the compression ratio for analysis.
We decompress the compressed activation data in the backpropagation when needed. As discussed in Section \ref{sec:analysis}, we need to force zeros in activation data to remain unchanged in compression algorithm. 
However, the current cuSZ algorithm may result in a series of small values within the error bound when encountering continuous zeros.  

To solve this issue, we propose a way to bypass this, which is to take advantage of the data recomputation technique.
Specifically, we recalculate the activation function layer when this happens, so that these zeros can remain zero after decompression, while the other values are still decompressed values.
For a more general solution, we propose to modify cuSZ for the case of compressing continuous zeros.
More specifically, we add a filter to the decompression process to re-zero those values within the error bound. 
Note that we propose to modify the decompression algorithm instead of the compression algorithm because it would not affect our compression ratio but only introduce little extra overhead. 

\section{Experimental Evaluation}
\label{sec:evaluation}

In this section, we evaluate our proposed framework from four aspects, including (1) evaluation of compression error impact on gradient, (2) evaluation of error propagation from gradient to training curve, (3) comparison between our proposed framework and the state-of-the-art method, and (4) performance evaluation.

\subsection{Experimental Setup}

Our evaluation are conducted with Caffe \cite{jia2014caffe} and Tensorflow \cite{abadi2016tensorflow}. We choose Caffe for single-node experiment due to its easy-to-modify and choose Tensorflow for multi-node evaluation due to its wise use. 
Our experiment platform is the TACC Longhorn system \cite{longhorn}, of which each GPU node is equipped with 4 Nvidia Tesla V100 GPUs \cite{nv100} per node. Our evaluation dataset is the ImageNet-2012 \cite{krizhevsky2012imagenet}. 
We use the CNN models for image classification including AlexNet \cite{krizhevsky2012imagenet}, VGG-16 \cite{simonyan2014very}, ResNet-18 and ResNet-50 \cite{he2016deep}.

\subsection{Error Propagation Evaluation}

\begin{figure}[]
    \centering
    \includegraphics[width=0.73\linewidth]{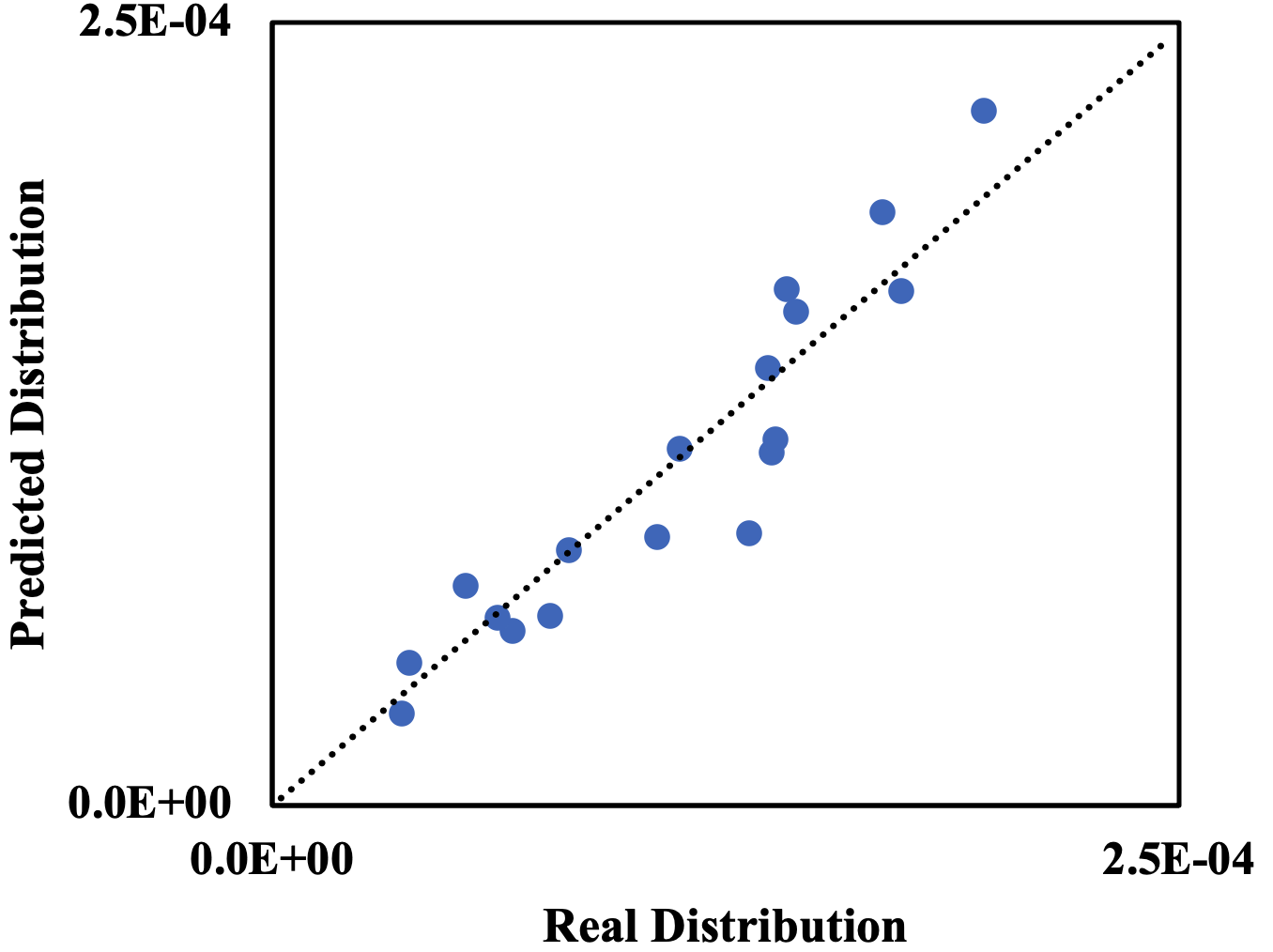}
    \vspace{-2mm}
    \caption{Comparison of $\sigma$ from real measured distribution and from predicted distribution.}
    \label{fig:fig-5-dis}
\end{figure}

First, we evaluate our proposed theoretical analysis in Section \ref{sub:theory-analysis}.
Based on Equation~\ref{equ-5} and Equation~\ref{equ-9}, we can estimate $\sigma$ which stands for how error is distributed in the gradient. 
After implementing our estimation, we identify that coefficient $a$ in Equation~\ref{equ-5} is $0.32$ based on our experiment. 
This is reasonable because if we consider the extreme condition that the batch size $N = 1$, the error distribution in the gradient will be the same as the SZ lossy compression to uniformly distribute and result in $a=1/3$. 
We also evaluate our estimation on different layers of AlexNet and VGG-16 using the batch size of 256, as shown in Figure~\ref{fig:fig-5-dis}. 
We can clearly observe that the coefficient and how our our estimated value aligns with the actual error distribution. 
This means that we can not only estimate the error propagation, but also determine the error bound based on a given acceptable $\sigma$ error distribution.

\begin{figure}[]
    \hspace*{-6mm}
    \includegraphics[width=0.92\linewidth]{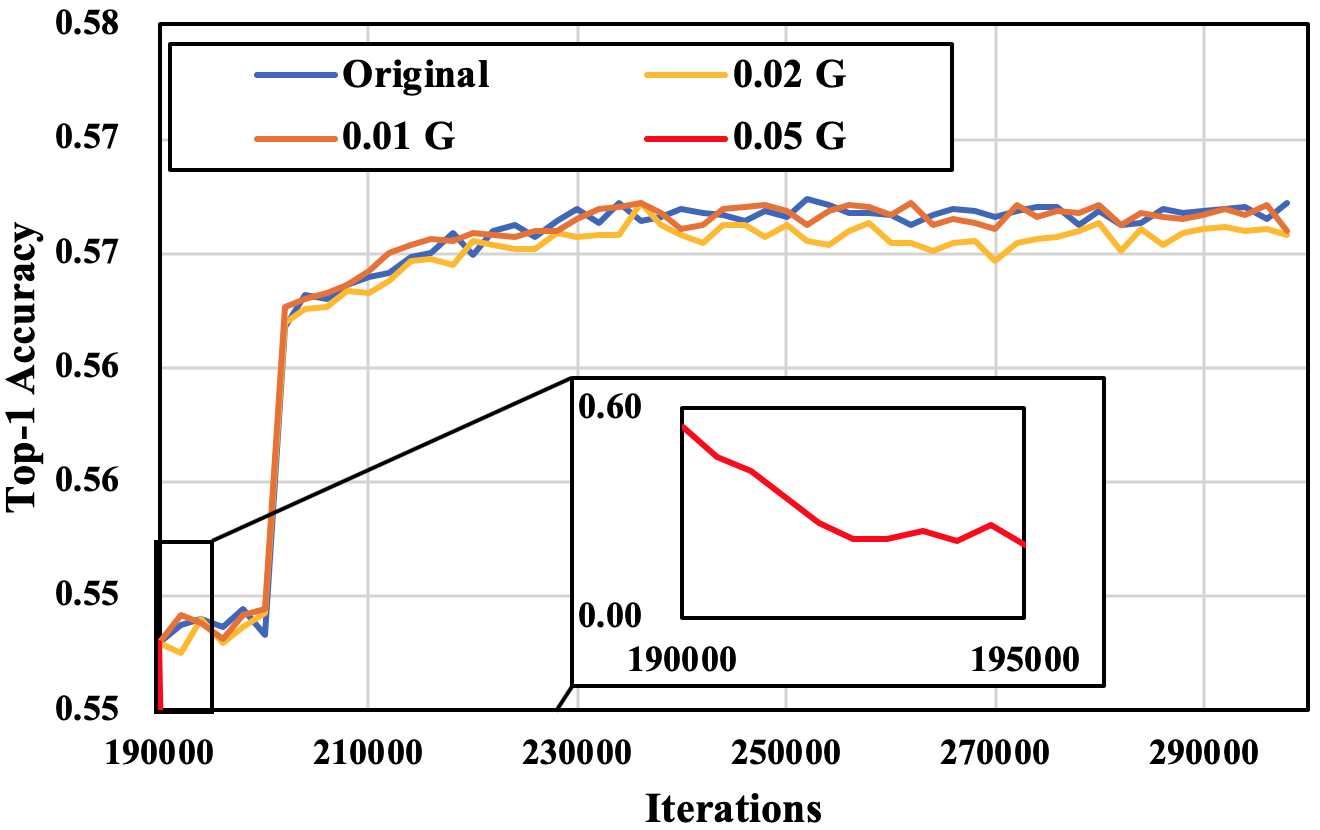}
    \vspace{-2mm}
    \caption{Evaluation on the overall training accuracy curve with different fraction of error introduced to the gradient. $G$ represents for the average value of gradient with the coefficient value used in our framework. }
    \label{fig:fig-5-sigma}
\end{figure}

Next, we evaluate the error propagation from gradient to overall training curve in terms of accuracy.
As we target to cause little or no accuracy loss, we focus on the iterations close the end of training in this evaluation, since the accuracy is hard to be increased when the training is close to the end. 
To reduce the training time and find an empirical solution, we pre-train the model without the help of our framework and save the snapshot every epoch. Then, we perform our evaluation of error propagation from different iterations from different snapshots. 
Figure~\ref{fig:fig-5-sigma} shows our experiment with AlexNet starting from the iteration of 190,000 with the batch size of 256.
On the one hand, we can observe from the zoomed in sub-figure that $\sigma = 0.05$ would result in an unacceptable error loss that cannot be eventually recovered. 
On the other hand, $\sigma = 0.02$ can provide better accuracy and higher compression ratio, but it does affect the accuracy a bit in some cases. 
Thus, considering that our goal is a general solution for all convolutional layers, we eventually choose $\sigma = 0.01$ as default in our framework; in other words, the target $\sigma$ is 1\% of the average of gradient. 

\subsection{Memory Reduction Evaluation}

\begin{figure}[]
    \centering
    \includegraphics[width=1.0\linewidth]{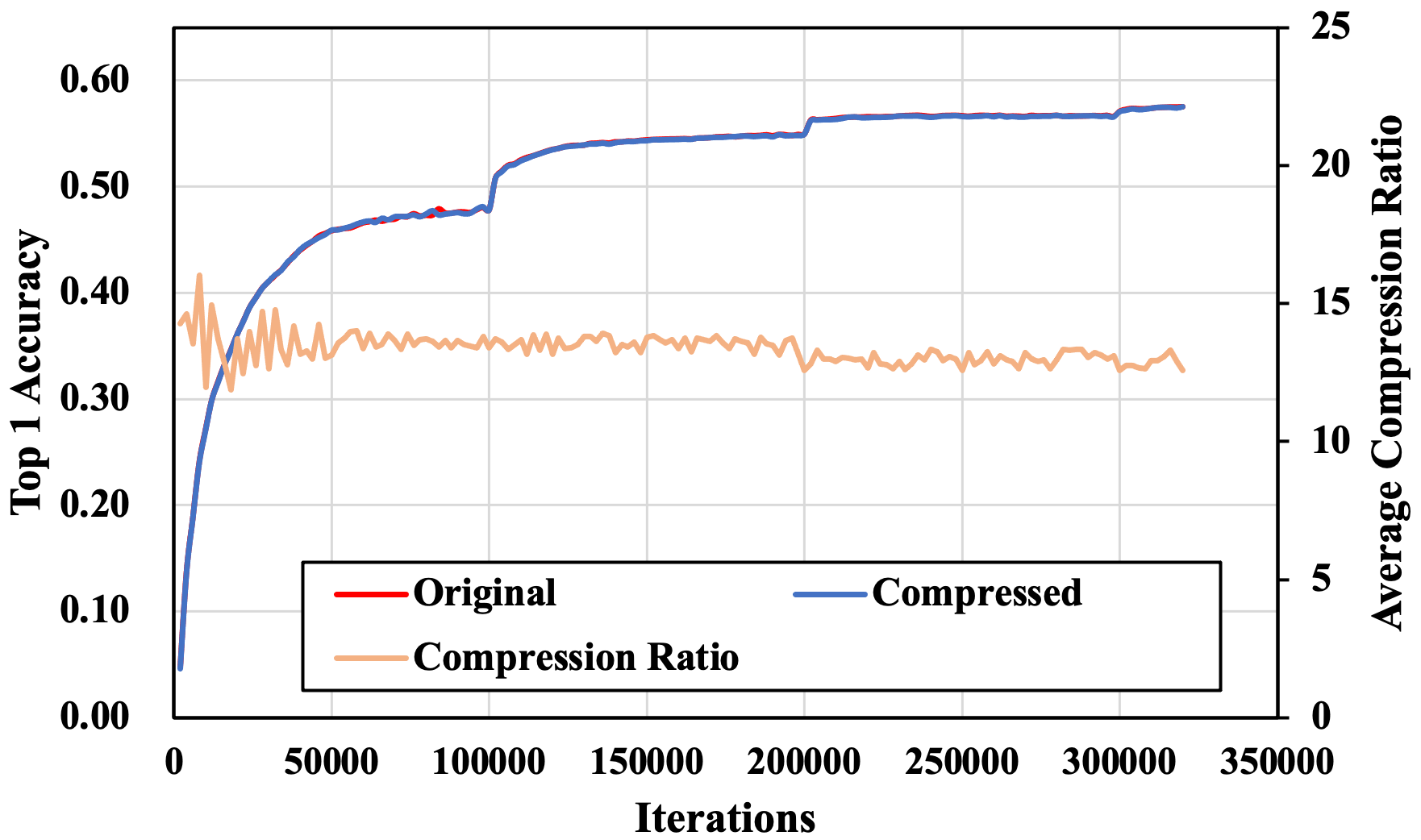}
    \caption{Training accuracy curve comparison between the baseline and our proposed framework (batch size = 256).}
    \vspace{-2mm}
    \label{fig:fig-5-alexnet}
\end{figure}

We then test our framework on various popular CNNs with the ImageNet dataset and evaluate its memory reduction capability. Figure~\ref{fig:fig-5-alexnet} illustrates the result with AlexNet. The red and blue lines are the training accuracy curves of baseline training and our framework. We can observe that these two curves are very close to each other, meaning our framework does not obviously affect the training accuracy. We also illustrate the curve of compression ratio to iteration in yellow in this figure. In the early stage of the training, compression ratio can be slightly unstable because of the relatively large change to the model. Note that the compression ratio will change slightly when the learning rate changes, because the learning rate only matters when updating the gradient to the weights.
Moreover, for some layers, although the average maximum loss of each input should be decreased and result in a higher error bound, the corresponding activation data value is actually increased. Thus, the compression ratio would not increase even with a higher error bound.

Table~\ref{tab:compress} shows the compression ratio of convolutional layers that our framework can provide. 
There is almost no accuracy loss or only little, with up to $0.31$\% .
This thanks to our careful control of compression error and thorough theoretical analysis and modeling of error impact. 
Our framework can deliver a promising compression ratio without heavy efforts of fine-tuning any parameter for different models. 
Overall, our proposed framework can provide up to 13.5$\times$ compression ratio with little or no accuracy loss. 

Compared with the lossless compression based solution \cite{rhu2018compressing}, which reduces the memory usage by within 2$\times$, our framework outperforms it by over 9$\times$; 
compared with the the current state-of-the-art lossy compression based solution \cite{evans2020jpeg}, which uses an image based lossy compressor to provide up to 7$\times$ compression ratios, our framework outperforms it by 1.5$\times$ and 1.8$\times$ on ResNet-18 and ResNet-50, respectively.

\begin{table}[t]
    \caption{Comparison of accuracy and activation size between baseline training and our proposed framework.}
    \label{tab:compress}
    \footnotesize{
\resizebox{\columnwidth}{!}{%
\begin{tabular}{@{}rrrr@{}}
\toprule
\textbf{Neural Network} & 
    \begin{tabular}{@{}r@{}} \bfseries Top-1 \\\bfseries Accuracy \end{tabular} & 
    \begin{tabular}{@{}r@{}} \bfseries Convolutional \\ \bfseries Act. Size \end{tabular} & 
    \begin{tabular}{@{}r@{}} \bfseries Compress \\\bfseries Ratio \end{tabular} \\
\midrule
baseline & 57.41\% & 407 \makebox[1.5em][l]{MB} & \\
\textbf{AlexNet} compressed & 57.42\% & \bfseries 30 \makebox[1.5em][l]{MB} & 13.5$\times$ \\
\cmidrule{2-4}
baseline & 
	68.05\% & 
	9.30 \makebox[1.5em][l]{GB} & \\
\textbf{VGG-16} compressed & 
	68.02\% & 
	\bfseries 0.83 \makebox[1.5em][l]{GB} & 
	11.1 $\times$ \\
\cmidrule{2-4}
baseline & 
	67.57\% & 
	3.42 \makebox[1.5em][l]{GB} & 
	\\
\textbf{ResNet-18} compressed & 
	67.43\% & 
	\bfseries 0.32 \makebox[1.5em][l]{GB} & 
	10.7 $\times$ \\
\cmidrule{2-4}
baseline & 
	71.49\% & 
	10.28 \makebox[1.5em][l]{GB} & 
	\\
\textbf{ResNet-50} compressed & 
	71.18\% & 
	\bfseries 0.93 \makebox[1.5em][l]{GB} & 
	11.0 $\times$ \\
\bottomrule
\end{tabular}
}}
\end{table}

\subsection{Performance Evaluation and Analysis}

\begin{figure}[]
    \centering
    \includegraphics[width=0.9\linewidth]{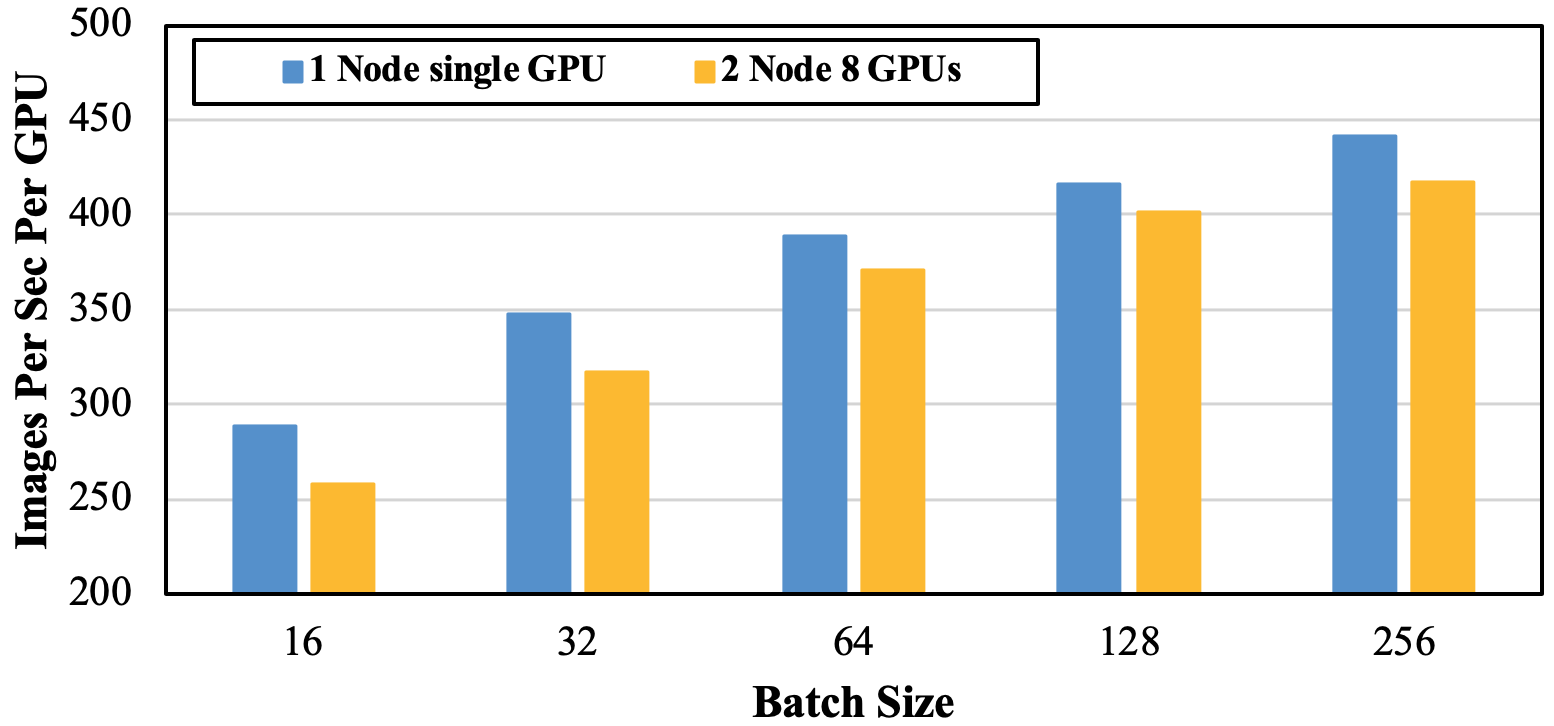}
    \vspace{-2mm}
    \caption{Training performance on ResNet-50 with different $N$.}
    \label{fig:fig-5-performance}
\end{figure}

Our framework introduces relatively small overhead to the training process while can greatly reduce the memory utilization and allow larger and wider neural networks to be trained with limited GPU memory. Moreover, the saved memory can also be further utilized for larger batch size, which improves the overall performance. Figure~\ref{fig:fig-5-performance} shows the improvement of raw performance (i.e., images per second) with increasing batch size on both single-GPU and multi-node/multi-GPU cases. We can clearly observe the raw performance can be improved with both use cases.

Another potential way to improve the performance from increasing the batch size is faster convergence speed to well trained status \cite{you2017scaling}. More batch size can lead to a more precise direction for the gradient instead of just rely on methods such as the momentum to reduce the impact of gradient uncertainty. Considering the optimized memory utilization from our framework, we can provide a up to 1.27$\times$ raw performance improvement, while providing the benefits from larger batch size.

Regarding the performance overhead of our framework, 
on the one hand, it needs to extract usable parameters and compute the compression configuration every 1000 iteration, while the amortized overhead is almost negligible;
on the other hand, thanks to its high working efficiency, cuSZ can provide an extremely high compression and decompression throughput on GPU \cite{tian2020cusz}. 
The only modification made to cuSZ is the filter that changes all the values under the error bound to zeros, as discussed in Section \ref{sub:comp}. 
This helps us to keep zeros in the activation data unchanged while only causing little overhead to the framework. 
Overall, our proposed framework introduces about 17\% overhead the training process while keeping the same training batch size. 
Moreover, our framework can further utilize the saved memory to increase the batch size and improve the training performance to offset this performance overhead. 
For example, our framework can achieve as low as 7\% overhead on VGG-16 by increasing the batch size from 32 to 256 with the similar memory consumption. 
In comparison, the state-of-the-art migration solution such as Layrub on average achieves a memory reduction of 2.4$\times$ but with a high training performance overhead of 24.1\% \cite{liu2018layrub}.
We also acknowledge that our framework may introduce high overheads to the networks that contain convolutional layers with many 1$\times$1 kernels. This is because 1$\times$1 kernels take little time to recompute but requires a relatively high overhead to compress and decompress. 
This is because calculating such layers is very efficient, compared with the GPU (de)compression on similar sizes of activation data.
Thus, our framework is more suitable for the CNNs composed of larger convolution kernels. 
\section{Conclusion and Future Work}
\label{sec:conclusion}

In this paper, we propose a novel memory-efficient deep learning training framework. 
We utilize the SZ error-bounded lossy compressor to reduce the memory consumption of convolutional layers. 
We develop an error propagation model and prove its accuracy. 
We evaluate our proposed framework on several popular CNNs with the ImageNet dataset.
The result shows that our framework significantly reduces the memory usage by up to 13.5$\times$ with little or no accuracy loss. 
Compared with the state-of-the-art compression based approach, our framework can provide an performance improvement of up to 1.8$\times$. 
We plan to implement those orthogonal methods such as data migration and recomputation into the framework for higher performance and more memory reduction.
Moreover, we will further optimize the compression and decompression performance on the activation data. 
\section*{Acknowledgments}
This research is supported by the National Science Foundation under Grants OAC-2034169 and OAC-2003624. The authors acknowledge the Texas Advanced Computing Center (TACC) at The University of Texas at Austin for providing HPC resources that have contributed to the research results reported within this paper.

\bibliographystyle{ACM-Reference-Format}
\bibliography{refs}


\begin{thebibliography}{52}


\ifx \showCODEN    \undefined \def \showCODEN     #1{\unskip}     \fi
\ifx \showDOI      \undefined \def \showDOI       #1{#1}\fi
\ifx \showISBNx    \undefined \def \showISBNx     #1{\unskip}     \fi
\ifx \showISBNxiii \undefined \def \showISBNxiii  #1{\unskip}     \fi
\ifx \showISSN     \undefined \def \showISSN      #1{\unskip}     \fi
\ifx \showLCCN     \undefined \def \showLCCN      #1{\unskip}     \fi
\ifx \shownote     \undefined \def \shownote      #1{#1}          \fi
\ifx \showarticletitle \undefined \def \showarticletitle #1{#1}   \fi
\ifx \showURL      \undefined \def \showURL       {\relax}        \fi
\providecommand\bibfield[2]{#2}
\providecommand\bibinfo[2]{#2}
\providecommand\natexlab[1]{#1}
\providecommand\showeprint[2][]{arXiv:#2}

\bibitem[\protect\citeauthoryear{Abadi, Agarwal, Barham, Brevdo, Chen, Citro,
  Corrado, Davis, Dean, Devin, et~al\mbox{.}}{Abadi et~al\mbox{.}}{2016}]%
        {abadi2016tensorflow}
\bibfield{author}{\bibinfo{person}{Mart{\'\i}n Abadi}, \bibinfo{person}{Ashish
  Agarwal}, \bibinfo{person}{Paul Barham}, \bibinfo{person}{Eugene Brevdo},
  \bibinfo{person}{Zhifeng Chen}, \bibinfo{person}{Craig Citro},
  \bibinfo{person}{Greg~S Corrado}, \bibinfo{person}{Andy Davis},
  \bibinfo{person}{Jeffrey Dean}, \bibinfo{person}{Matthieu Devin},
  {et~al\mbox{.}}} \bibinfo{year}{2016}\natexlab{}.
\newblock \showarticletitle{Tensorflow: Large-scale machine learning on
  heterogeneous distributed systems}.
\newblock \bibinfo{journal}{\emph{arXiv preprint arXiv:1603.04467}}
  (\bibinfo{year}{2016}).
\newblock


\bibitem[\protect\citeauthoryear{Ben-Nun and Hoefler}{Ben-Nun and
  Hoefler}{2019}]%
        {ben2019demystifying}
\bibfield{author}{\bibinfo{person}{Tal Ben-Nun} {and} \bibinfo{person}{Torsten
  Hoefler}.} \bibinfo{year}{2019}\natexlab{}.
\newblock \showarticletitle{Demystifying parallel and distributed deep
  learning: An in-depth concurrency analysis}.
\newblock \bibinfo{journal}{\emph{ACM Computing Surveys (CSUR)}}
  \bibinfo{volume}{52}, \bibinfo{number}{4} (\bibinfo{year}{2019}),
  \bibinfo{pages}{1--43}.
\newblock


\bibitem[\protect\citeauthoryear{Brown, Mann, Ryder, Subbiah, Kaplan, Dhariwal,
  Neelakantan, Shyam, Sastry, Askell, et~al\mbox{.}}{Brown
  et~al\mbox{.}}{2020}]%
        {brown2020language}
\bibfield{author}{\bibinfo{person}{Tom~B Brown}, \bibinfo{person}{Benjamin
  Mann}, \bibinfo{person}{Nick Ryder}, \bibinfo{person}{Melanie Subbiah},
  \bibinfo{person}{Jared Kaplan}, \bibinfo{person}{Prafulla Dhariwal},
  \bibinfo{person}{Arvind Neelakantan}, \bibinfo{person}{Pranav Shyam},
  \bibinfo{person}{Girish Sastry}, \bibinfo{person}{Amanda Askell},
  {et~al\mbox{.}}} \bibinfo{year}{2020}\natexlab{}.
\newblock \showarticletitle{Language models are few-shot learners}.
\newblock \bibinfo{journal}{\emph{arXiv preprint arXiv:2005.14165}}
  (\bibinfo{year}{2020}).
\newblock


\bibitem[\protect\citeauthoryear{Burtscher and Ratanaworabhan}{Burtscher and
  Ratanaworabhan}{2008}]%
        {FPC}
\bibfield{author}{\bibinfo{person}{Martin Burtscher} {and}
  \bibinfo{person}{Paruj Ratanaworabhan}.} \bibinfo{year}{2008}\natexlab{}.
\newblock \showarticletitle{FPC: A high-speed compressor for double-precision
  floating-point data}.
\newblock \bibinfo{journal}{\emph{IEEE Trans. Comput.}} \bibinfo{volume}{58},
  \bibinfo{number}{1} (\bibinfo{year}{2008}), \bibinfo{pages}{18--31}.
\newblock


\bibitem[\protect\citeauthoryear{Challenge}{Challenge}{2020}]%
        {ILSVRC}
\bibfield{author}{\bibinfo{person}{Large Scale Visual~Recognition Challenge}.}
  \bibinfo{year}{2020}\natexlab{}.
\newblock
  \bibinfo{howpublished}{\url{http://www.image-net.org/challenges/LSVRC/}}.
\newblock


\bibitem[\protect\citeauthoryear{Chen, Xu, Zhang, and Guestrin}{Chen
  et~al\mbox{.}}{2016}]%
        {chen2016training}
\bibfield{author}{\bibinfo{person}{Tianqi Chen}, \bibinfo{person}{Bing Xu},
  \bibinfo{person}{Chiyuan Zhang}, {and} \bibinfo{person}{Carlos Guestrin}.}
  \bibinfo{year}{2016}\natexlab{}.
\newblock \showarticletitle{Training deep nets with sublinear memory cost}.
\newblock \bibinfo{journal}{\emph{arXiv preprint arXiv:1604.06174}}
  (\bibinfo{year}{2016}).
\newblock


\bibitem[\protect\citeauthoryear{Choukse, Sullivan, O’Connor, Erez, Pool,
  Nellans, and Keckler}{Choukse et~al\mbox{.}}{2020}]%
        {choukse2020buddy}
\bibfield{author}{\bibinfo{person}{Esha Choukse}, \bibinfo{person}{Michael~B
  Sullivan}, \bibinfo{person}{Mike O’Connor}, \bibinfo{person}{Mattan Erez},
  \bibinfo{person}{Jeff Pool}, \bibinfo{person}{David Nellans}, {and}
  \bibinfo{person}{Stephen~W Keckler}.} \bibinfo{year}{2020}\natexlab{}.
\newblock \showarticletitle{Buddy compression: Enabling larger memory for deep
  learning and HPC workloads on gpus}. In \bibinfo{booktitle}{\emph{2020
  ACM/IEEE 47th Annual International Symposium on Computer Architecture
  (ISCA)}}. IEEE, \bibinfo{pages}{926--939}.
\newblock


\bibitem[\protect\citeauthoryear{Collobert and Weston}{Collobert and
  Weston}{2008}]%
        {collobert2008unified}
\bibfield{author}{\bibinfo{person}{Ronan Collobert} {and}
  \bibinfo{person}{Jason Weston}.} \bibinfo{year}{2008}\natexlab{}.
\newblock \showarticletitle{A unified architecture for natural language
  processing: Deep neural networks with multitask learning}. In
  \bibinfo{booktitle}{\emph{Proceedings of the 25th International Conference on
  Machine Learning}}. ACM, \bibinfo{pages}{160--167}.
\newblock


\bibitem[\protect\citeauthoryear{Cui, Zhang, Ganger, Gibbons, and Xing}{Cui
  et~al\mbox{.}}{2016}]%
        {cui2016geeps}
\bibfield{author}{\bibinfo{person}{Henggang Cui}, \bibinfo{person}{Hao Zhang},
  \bibinfo{person}{Gregory~R Ganger}, \bibinfo{person}{Phillip~B Gibbons},
  {and} \bibinfo{person}{Eric~P Xing}.} \bibinfo{year}{2016}\natexlab{}.
\newblock \showarticletitle{Geeps: Scalable deep learning on distributed gpus
  with a gpu-specialized parameter server}. In
  \bibinfo{booktitle}{\emph{Proceedings of the Eleventh European Conference on
  Computer Systems}}. ACM, \bibinfo{pages}{4}.
\newblock


\bibitem[\protect\citeauthoryear{Deutsch}{Deutsch}{1996}]%
        {gzip}
\bibfield{author}{\bibinfo{person}{Peter Deutsch}.}
  \bibinfo{year}{1996}\natexlab{}.
\newblock \bibinfo{booktitle}{\emph{GZIP file format specification version
  4.3}}.
\newblock \bibinfo{type}{{T}echnical {R}eport}.
\newblock


\bibitem[\protect\citeauthoryear{Di and Cappello}{Di and Cappello}{2016}]%
        {di2016fast}
\bibfield{author}{\bibinfo{person}{Sheng Di} {and} \bibinfo{person}{Franck
  Cappello}.} \bibinfo{year}{2016}\natexlab{}.
\newblock \showarticletitle{Fast error-bounded lossy {HPC} data compression
  with {SZ}}. In \bibinfo{booktitle}{\emph{2016 IEEE International Parallel and
  Distributed Processing Symposium}}. IEEE, \bibinfo{pages}{730--739}.
\newblock


\bibitem[\protect\citeauthoryear{Dong, Liao, Pang, Su, Zhu, Hu, and Li}{Dong
  et~al\mbox{.}}{2018}]%
        {dong2018boosting}
\bibfield{author}{\bibinfo{person}{Yinpeng Dong}, \bibinfo{person}{Fangzhou
  Liao}, \bibinfo{person}{Tianyu Pang}, \bibinfo{person}{Hang Su},
  \bibinfo{person}{Jun Zhu}, \bibinfo{person}{Xiaolin Hu}, {and}
  \bibinfo{person}{Jianguo Li}.} \bibinfo{year}{2018}\natexlab{}.
\newblock \showarticletitle{Boosting adversarial attacks with momentum}. In
  \bibinfo{booktitle}{\emph{Proceedings of the IEEE conference on computer
  vision and pattern recognition}}. \bibinfo{pages}{9185--9193}.
\newblock


\bibitem[\protect\citeauthoryear{Ekman and Stenstrom}{Ekman and
  Stenstrom}{2005}]%
        {ekman2005robust}
\bibfield{author}{\bibinfo{person}{Magnus Ekman} {and} \bibinfo{person}{Per
  Stenstrom}.} \bibinfo{year}{2005}\natexlab{}.
\newblock \showarticletitle{A robust main-memory compression scheme}. In
  \bibinfo{booktitle}{\emph{32nd International Symposium on Computer
  Architecture (ISCA'05)}}. IEEE, \bibinfo{pages}{74--85}.
\newblock


\bibitem[\protect\citeauthoryear{Evans, Liu, and Aamodt}{Evans
  et~al\mbox{.}}{2020}]%
        {evans2020jpeg}
\bibfield{author}{\bibinfo{person}{R~David Evans}, \bibinfo{person}{Lufei Liu},
  {and} \bibinfo{person}{Tor~M Aamodt}.} \bibinfo{year}{2020}\natexlab{}.
\newblock \showarticletitle{JPEG-ACT: Accelerating Deep Learning via
  Transform-based Lossy Compression}. In \bibinfo{booktitle}{\emph{2020
  ACM/IEEE 47th Annual International Symposium on Computer Architecture
  (ISCA)}}. IEEE, \bibinfo{pages}{860--873}.
\newblock


\bibitem[\protect\citeauthoryear{Foley and Danskin}{Foley and Danskin}{2017}]%
        {foley2017ultra}
\bibfield{author}{\bibinfo{person}{Denis Foley} {and} \bibinfo{person}{John
  Danskin}.} \bibinfo{year}{2017}\natexlab{}.
\newblock \showarticletitle{Ultra-performance Pascal GPU and NVLink
  interconnect}.
\newblock \bibinfo{journal}{\emph{IEEE Micro}} \bibinfo{volume}{37},
  \bibinfo{number}{2} (\bibinfo{year}{2017}), \bibinfo{pages}{7--17}.
\newblock


\bibitem[\protect\citeauthoryear{Gomez, Ren, Urtasun, and Grosse}{Gomez
  et~al\mbox{.}}{2017}]%
        {gomez2017reversible}
\bibfield{author}{\bibinfo{person}{Aidan~N Gomez}, \bibinfo{person}{Mengye
  Ren}, \bibinfo{person}{Raquel Urtasun}, {and} \bibinfo{person}{Roger~B
  Grosse}.} \bibinfo{year}{2017}\natexlab{}.
\newblock \showarticletitle{The reversible residual network: Backpropagation
  without storing activations}. In \bibinfo{booktitle}{\emph{Advances in neural
  information processing systems}}. \bibinfo{pages}{2214--2224}.
\newblock


\bibitem[\protect\citeauthoryear{Google}{Google}{2018}]%
        {googletpu}
\bibfield{author}{\bibinfo{person}{Google}.} \bibinfo{year}{2018}\natexlab{}.
\newblock \bibinfo{howpublished}{\url{https://cloud.google.com/tpu}}.
\newblock


\bibitem[\protect\citeauthoryear{GPU}{GPU}{2020}]%
        {nv100}
\bibfield{author}{\bibinfo{person}{NVIDIA V100 TENSOR~CORE GPU}.}
  \bibinfo{year}{2020}\natexlab{}.
\newblock
  \bibinfo{howpublished}{\url{https://www.nvidia.com/en-us/data-center/v100/}}.
\newblock


\bibitem[\protect\citeauthoryear{He, Zhang, Ren, and Sun}{He
  et~al\mbox{.}}{2016}]%
        {he2016deep}
\bibfield{author}{\bibinfo{person}{Kaiming He}, \bibinfo{person}{Xiangyu
  Zhang}, \bibinfo{person}{Shaoqing Ren}, {and} \bibinfo{person}{Jian Sun}.}
  \bibinfo{year}{2016}\natexlab{}.
\newblock \showarticletitle{Deep residual learning for image recognition}. In
  \bibinfo{booktitle}{\emph{Proceedings of the IEEE conference on Computer
  Vision and Pattern Recognition}}. \bibinfo{pages}{770--778}.
\newblock


\bibitem[\protect\citeauthoryear{Hecht-Nielsen}{Hecht-Nielsen}{1992}]%
        {hecht1992theory}
\bibfield{author}{\bibinfo{person}{Robert Hecht-Nielsen}.}
  \bibinfo{year}{1992}\natexlab{}.
\newblock \showarticletitle{Theory of the backpropagation neural network}.
\newblock In \bibinfo{booktitle}{\emph{Neural networks for perception}}.
  \bibinfo{publisher}{Elsevier}, \bibinfo{pages}{65--93}.
\newblock


\bibitem[\protect\citeauthoryear{Huang, Cheng, Bapna, Firat, Chen, Chen, Lee,
  Ngiam, Le, Wu, et~al\mbox{.}}{Huang et~al\mbox{.}}{2019}]%
        {huang2019gpipe}
\bibfield{author}{\bibinfo{person}{Yanping Huang}, \bibinfo{person}{Youlong
  Cheng}, \bibinfo{person}{Ankur Bapna}, \bibinfo{person}{Orhan Firat},
  \bibinfo{person}{Dehao Chen}, \bibinfo{person}{Mia Chen},
  \bibinfo{person}{HyoukJoong Lee}, \bibinfo{person}{Jiquan Ngiam},
  \bibinfo{person}{Quoc~V Le}, \bibinfo{person}{Yonghui Wu}, {et~al\mbox{.}}}
  \bibinfo{year}{2019}\natexlab{}.
\newblock \showarticletitle{Gpipe: Efficient training of giant neural networks
  using pipeline parallelism}. In \bibinfo{booktitle}{\emph{Advances in neural
  information processing systems}}. \bibinfo{pages}{103--112}.
\newblock


\bibitem[\protect\citeauthoryear{Jia, Shelhamer, Donahue, Karayev, Long,
  Girshick, Guadarrama, and Darrell}{Jia et~al\mbox{.}}{2014}]%
        {jia2014caffe}
\bibfield{author}{\bibinfo{person}{Yangqing Jia}, \bibinfo{person}{Evan
  Shelhamer}, \bibinfo{person}{Jeff Donahue}, \bibinfo{person}{Sergey Karayev},
  \bibinfo{person}{Jonathan Long}, \bibinfo{person}{Ross Girshick},
  \bibinfo{person}{Sergio Guadarrama}, {and} \bibinfo{person}{Trevor Darrell}.}
  \bibinfo{year}{2014}\natexlab{}.
\newblock \showarticletitle{Caffe: Convolutional architecture for fast feature
  embedding}. In \bibinfo{booktitle}{\emph{Proceedings of the 22nd ACM
  international conference on Multimedia}}. ACM, \bibinfo{pages}{675--678}.
\newblock


\bibitem[\protect\citeauthoryear{Jin, Di, Liang, Tian, Tao, and Cappello}{Jin
  et~al\mbox{.}}{2019}]%
        {jin2019deepsz}
\bibfield{author}{\bibinfo{person}{Sian Jin}, \bibinfo{person}{Sheng Di},
  \bibinfo{person}{Xin Liang}, \bibinfo{person}{Jiannan Tian},
  \bibinfo{person}{Dingwen Tao}, {and} \bibinfo{person}{Franck Cappello}.}
  \bibinfo{year}{2019}\natexlab{}.
\newblock \showarticletitle{Deepsz: A novel framework to compress deep neural
  networks by using error-bounded lossy compression}. In
  \bibinfo{booktitle}{\emph{Proceedings of the 28th International Symposium on
  High-Performance Parallel and Distributed Computing}}.
  \bibinfo{pages}{159--170}.
\newblock


\bibitem[\protect\citeauthoryear{Krizhevsky, Sutskever, and Hinton}{Krizhevsky
  et~al\mbox{.}}{2012}]%
        {krizhevsky2012imagenet}
\bibfield{author}{\bibinfo{person}{Alex Krizhevsky}, \bibinfo{person}{Ilya
  Sutskever}, {and} \bibinfo{person}{Geoffrey~E Hinton}.}
  \bibinfo{year}{2012}\natexlab{}.
\newblock \showarticletitle{Imagenet classification with deep convolutional
  neural networks}. In \bibinfo{booktitle}{\emph{Advances in neural information
  processing systems}}. \bibinfo{pages}{1097--1105}.
\newblock


\bibitem[\protect\citeauthoryear{Lal, Lucas, and Juurlink}{Lal
  et~al\mbox{.}}{2017}]%
        {lal20172mc}
\bibfield{author}{\bibinfo{person}{Sohan Lal}, \bibinfo{person}{Jan Lucas},
  {and} \bibinfo{person}{Ben Juurlink}.} \bibinfo{year}{2017}\natexlab{}.
\newblock \showarticletitle{E\^{} 2MC: Entropy Encoding Based Memory
  Compression for GPUs}. In \bibinfo{booktitle}{\emph{2017 IEEE International
  Parallel and Distributed Processing Symposium (IPDPS)}}. IEEE,
  \bibinfo{pages}{1119--1128}.
\newblock


\bibitem[\protect\citeauthoryear{Li}{Li}{2020}]%
        {gtp-price}
\bibfield{author}{\bibinfo{person}{Chuan Li}.} \bibinfo{year}{2020}\natexlab{}.
\newblock \bibinfo{title}{{OpenAI's GPT-3 Language Model: A Technical
  Overview}}.
\newblock
  \bibinfo{howpublished}{\url{https://lambdalabs.com/blog/demystifying-gpt-3/}}.
\newblock


\bibitem[\protect\citeauthoryear{Liang, Di, Tao, Li, Li, Guo, Chen, and
  Cappello}{Liang et~al\mbox{.}}{2018}]%
        {liangerror}
\bibfield{author}{\bibinfo{person}{Xin Liang}, \bibinfo{person}{Sheng Di},
  \bibinfo{person}{Dingwen Tao}, \bibinfo{person}{Sihuan Li},
  \bibinfo{person}{Shaomeng Li}, \bibinfo{person}{Hanqi Guo},
  \bibinfo{person}{Zizhong Chen}, {and} \bibinfo{person}{Franck Cappello}.}
  \bibinfo{year}{2018}\natexlab{}.
\newblock \showarticletitle{Error-Controlled Lossy Compression Optimized for
  High Compression Ratios of Scientific Datasets}.
\newblock  (\bibinfo{year}{2018}).
\newblock


\bibitem[\protect\citeauthoryear{Lindstrom}{Lindstrom}{2014}]%
        {zfp}
\bibfield{author}{\bibinfo{person}{Peter Lindstrom}.}
  \bibinfo{year}{2014}\natexlab{}.
\newblock \showarticletitle{Fixed-rate compressed floating-point arrays}.
\newblock \bibinfo{journal}{\emph{IEEE Transactions on Visualization and
  Computer Graphics}} \bibinfo{volume}{20}, \bibinfo{number}{12}
  (\bibinfo{year}{2014}), \bibinfo{pages}{2674--2683}.
\newblock


\bibitem[\protect\citeauthoryear{Lindstrom}{Lindstrom}{2017}]%
        {lindstrom2017error}
\bibfield{author}{\bibinfo{person}{Peter Lindstrom}.}
  \bibinfo{year}{2017}\natexlab{}.
\newblock \bibinfo{booktitle}{\emph{Error distributions of lossy floating-point
  compressors}}.
\newblock \bibinfo{type}{{T}echnical {R}eport}. \bibinfo{institution}{Lawrence
  Livermore National Lab.(LLNL), Livermore, CA (United States)}.
\newblock


\bibitem[\protect\citeauthoryear{Lindstrom and Isenburg}{Lindstrom and
  Isenburg}{2006}]%
        {lindstrom2006fast}
\bibfield{author}{\bibinfo{person}{Peter Lindstrom} {and}
  \bibinfo{person}{Martin Isenburg}.} \bibinfo{year}{2006}\natexlab{}.
\newblock \showarticletitle{Fast and efficient compression of floating-point
  data}.
\newblock \bibinfo{journal}{\emph{IEEE Transactions on Visualization and
  Computer Graphics}} \bibinfo{volume}{12}, \bibinfo{number}{5}
  (\bibinfo{year}{2006}), \bibinfo{pages}{1245--1250}.
\newblock


\bibitem[\protect\citeauthoryear{Liu, Jiang, Jin, Shi, and Ma}{Liu
  et~al\mbox{.}}{2018}]%
        {liu2018layrub}
\bibfield{author}{\bibinfo{person}{Bo Liu}, \bibinfo{person}{Wenbin Jiang},
  \bibinfo{person}{Hai Jin}, \bibinfo{person}{Xuanhua Shi}, {and}
  \bibinfo{person}{Yang Ma}.} \bibinfo{year}{2018}\natexlab{}.
\newblock \showarticletitle{Layrub: layer-centric GPU memory reuse and data
  migration in extreme-scale deep learning systems}. In
  \bibinfo{booktitle}{\emph{Proceedings of the 23rd ACM SIGPLAN Symposium on
  Principles and Practice of Parallel Programming}}. \bibinfo{pages}{405--406}.
\newblock


\bibitem[\protect\citeauthoryear{{Longhorn subsystem}}{{Longhorn
  subsystem}}{2020}]%
        {longhorn}
\bibfield{author}{\bibinfo{person}{{Longhorn subsystem}}.}
  \bibinfo{year}{2020}\natexlab{}.
\newblock
  \bibinfo{howpublished}{\url{https://www.tacc.utexas.edu/systems/longhorn}}.
\newblock


\bibitem[\protect\citeauthoryear{Paszke, Gross, Massa, Lerer, Bradbury, Chanan,
  Killeen, Lin, Gimelshein, Antiga, et~al\mbox{.}}{Paszke
  et~al\mbox{.}}{2019}]%
        {paszke2019pytorch}
\bibfield{author}{\bibinfo{person}{Adam Paszke}, \bibinfo{person}{Sam Gross},
  \bibinfo{person}{Francisco Massa}, \bibinfo{person}{Adam Lerer},
  \bibinfo{person}{James Bradbury}, \bibinfo{person}{Gregory Chanan},
  \bibinfo{person}{Trevor Killeen}, \bibinfo{person}{Zeming Lin},
  \bibinfo{person}{Natalia Gimelshein}, \bibinfo{person}{Luca Antiga},
  {et~al\mbox{.}}} \bibinfo{year}{2019}\natexlab{}.
\newblock \showarticletitle{Pytorch: An imperative style, high-performance deep
  learning library}. In \bibinfo{booktitle}{\emph{Advances in neural
  information processing systems}}. \bibinfo{pages}{8026--8037}.
\newblock


\bibitem[\protect\citeauthoryear{Rhu, Gimelshein, Clemons, Zulfiqar, and
  Keckler}{Rhu et~al\mbox{.}}{2016}]%
        {rhu2016vdnn}
\bibfield{author}{\bibinfo{person}{Minsoo Rhu}, \bibinfo{person}{Natalia
  Gimelshein}, \bibinfo{person}{Jason Clemons}, \bibinfo{person}{Arslan
  Zulfiqar}, {and} \bibinfo{person}{Stephen~W Keckler}.}
  \bibinfo{year}{2016}\natexlab{}.
\newblock \showarticletitle{vDNN: Virtualized deep neural networks for
  scalable, memory-efficient neural network design}. In
  \bibinfo{booktitle}{\emph{The 49th Annual IEEE/ACM International Symposium on
  Microarchitecture}}. IEEE Press, \bibinfo{pages}{18}.
\newblock


\bibitem[\protect\citeauthoryear{Rhu, O'Connor, Chatterjee, Pool, Kwon, and
  Keckler}{Rhu et~al\mbox{.}}{2018}]%
        {rhu2018compressing}
\bibfield{author}{\bibinfo{person}{Minsoo Rhu}, \bibinfo{person}{Mike
  O'Connor}, \bibinfo{person}{Niladrish Chatterjee}, \bibinfo{person}{Jeff
  Pool}, \bibinfo{person}{Youngeun Kwon}, {and} \bibinfo{person}{Stephen~W
  Keckler}.} \bibinfo{year}{2018}\natexlab{}.
\newblock \showarticletitle{Compressing DMA engine: Leveraging activation
  sparsity for training deep neural networks}. In
  \bibinfo{booktitle}{\emph{2018 IEEE International Symposium on High
  Performance Computer Architecture (HPCA)}}. IEEE, \bibinfo{pages}{78--91}.
\newblock


\bibitem[\protect\citeauthoryear{Russakovsky, Deng, Su, Krause, Satheesh, Ma,
  Huang, Karpathy, Khosla, Bernstein, et~al\mbox{.}}{Russakovsky
  et~al\mbox{.}}{2015}]%
        {russakovsky2015imagenet}
\bibfield{author}{\bibinfo{person}{Olga Russakovsky}, \bibinfo{person}{Jia
  Deng}, \bibinfo{person}{Hao Su}, \bibinfo{person}{Jonathan Krause},
  \bibinfo{person}{Sanjeev Satheesh}, \bibinfo{person}{Sean Ma},
  \bibinfo{person}{Zhiheng Huang}, \bibinfo{person}{Andrej Karpathy},
  \bibinfo{person}{Aditya Khosla}, \bibinfo{person}{Michael Bernstein},
  {et~al\mbox{.}}} \bibinfo{year}{2015}\natexlab{}.
\newblock \showarticletitle{Imagenet large scale visual recognition challenge}.
\newblock \bibinfo{journal}{\emph{International journal of computer vision}}
  \bibinfo{volume}{115}, \bibinfo{number}{3} (\bibinfo{year}{2015}),
  \bibinfo{pages}{211--252}.
\newblock


\bibitem[\protect\citeauthoryear{Sergeev and Del~Balso}{Sergeev and
  Del~Balso}{2018}]%
        {sergeev2018horovod}
\bibfield{author}{\bibinfo{person}{Alexander Sergeev} {and}
  \bibinfo{person}{Mike Del~Balso}.} \bibinfo{year}{2018}\natexlab{}.
\newblock \showarticletitle{Horovod: fast and easy distributed deep learning in
  TensorFlow}.
\newblock \bibinfo{journal}{\emph{arXiv preprint arXiv:1802.05799}}
  (\bibinfo{year}{2018}).
\newblock


\bibitem[\protect\citeauthoryear{Simonyan and Zisserman}{Simonyan and
  Zisserman}{2014}]%
        {simonyan2014very}
\bibfield{author}{\bibinfo{person}{Karen Simonyan} {and}
  \bibinfo{person}{Andrew Zisserman}.} \bibinfo{year}{2014}\natexlab{}.
\newblock \showarticletitle{Very deep convolutional networks for large-scale
  image recognition}.
\newblock \bibinfo{journal}{\emph{arXiv preprint arXiv:1409.1556}}
  (\bibinfo{year}{2014}).
\newblock


\bibitem[\protect\citeauthoryear{Son, Chen, Hendrix, Agrawal, Liao, and
  Choudhary}{Son et~al\mbox{.}}{2014}]%
        {son2014data}
\bibfield{author}{\bibinfo{person}{Seung~Woo Son}, \bibinfo{person}{Zhengzhang
  Chen}, \bibinfo{person}{William Hendrix}, \bibinfo{person}{Ankit Agrawal},
  \bibinfo{person}{Wei-keng Liao}, {and} \bibinfo{person}{Alok Choudhary}.}
  \bibinfo{year}{2014}\natexlab{}.
\newblock \showarticletitle{Data compression for the exascale computing
  era-survey}.
\newblock \bibinfo{journal}{\emph{Supercomputing Frontiers and Innovations}}
  \bibinfo{volume}{1}, \bibinfo{number}{2} (\bibinfo{year}{2014}),
  \bibinfo{pages}{76--88}.
\newblock


\bibitem[\protect\citeauthoryear{{Sum of normally distributed random
  variables}}{{Sum of normally distributed random variables}}{[n. d.]}]%
        {wiki}
\bibfield{author}{\bibinfo{person}{{Sum of normally distributed random
  variables}}.} \bibinfo{year}{[n. d.]}\natexlab{}.
\newblock
  \bibinfo{howpublished}{\url{https://en.wikipedia.org/wiki/Sum_of_normally_distributed_random_variables}}.
\newblock


\bibitem[\protect\citeauthoryear{Szegedy, Ioffe, Vanhoucke, and Alemi}{Szegedy
  et~al\mbox{.}}{2017}]%
        {szegedy2017inception}
\bibfield{author}{\bibinfo{person}{Christian Szegedy}, \bibinfo{person}{Sergey
  Ioffe}, \bibinfo{person}{Vincent Vanhoucke}, {and}
  \bibinfo{person}{Alexander~A Alemi}.} \bibinfo{year}{2017}\natexlab{}.
\newblock \showarticletitle{Inception-v4, inception-resnet and the impact of
  residual connections on learning}. In \bibinfo{booktitle}{\emph{Thirty-first
  AAAI conference on artificial intelligence}}.
\newblock


\bibitem[\protect\citeauthoryear{Szegedy, Liu, Jia, Sermanet, Reed, Anguelov,
  Erhan, Vanhoucke, and Rabinovich}{Szegedy et~al\mbox{.}}{2015}]%
        {szegedy2015going}
\bibfield{author}{\bibinfo{person}{Christian Szegedy}, \bibinfo{person}{Wei
  Liu}, \bibinfo{person}{Yangqing Jia}, \bibinfo{person}{Pierre Sermanet},
  \bibinfo{person}{Scott Reed}, \bibinfo{person}{Dragomir Anguelov},
  \bibinfo{person}{Dumitru Erhan}, \bibinfo{person}{Vincent Vanhoucke}, {and}
  \bibinfo{person}{Andrew Rabinovich}.} \bibinfo{year}{2015}\natexlab{}.
\newblock \showarticletitle{Going deeper with convolutions}. In
  \bibinfo{booktitle}{\emph{Proceedings of the IEEE conference on Computer
  Vision and Pattern Recognition}}. \bibinfo{pages}{1--9}.
\newblock


\bibitem[\protect\citeauthoryear{Tao, Di, Chen, and Cappello}{Tao
  et~al\mbox{.}}{2017}]%
        {tao2017significantly}
\bibfield{author}{\bibinfo{person}{Dingwen Tao}, \bibinfo{person}{Sheng Di},
  \bibinfo{person}{Zizhong Chen}, {and} \bibinfo{person}{Franck Cappello}.}
  \bibinfo{year}{2017}\natexlab{}.
\newblock \showarticletitle{Significantly improving lossy compression for
  scientific data sets based on multidimensional prediction and
  error-controlled quantization}. In \bibinfo{booktitle}{\emph{2017 IEEE
  International Parallel and Distributed Processing Symposium}}. IEEE,
  \bibinfo{pages}{1129--1139}.
\newblock


\bibitem[\protect\citeauthoryear{Taubman and Marcellin}{Taubman and
  Marcellin}{2012}]%
        {taubman2012jpeg2000}
\bibfield{author}{\bibinfo{person}{David Taubman} {and}
  \bibinfo{person}{Michael Marcellin}.} \bibinfo{year}{2012}\natexlab{}.
\newblock \bibinfo{booktitle}{\emph{{JPEG2000} image compression fundamentals,
  standards and practice: image compression fundamentals, standards and
  practice}}. Vol.~\bibinfo{volume}{642}.
\newblock \bibinfo{publisher}{Springer Science \& Business Media}.
\newblock


\bibitem[\protect\citeauthoryear{Tian, Di, Zhao, Rivera, Fulp, Underwood, Jin,
  Liang, Calhoun, Tao, and Cappello}{Tian et~al\mbox{.}}{2020}]%
        {tian2020cusz}
\bibfield{author}{\bibinfo{person}{Jiannan Tian}, \bibinfo{person}{Sheng Di},
  \bibinfo{person}{Kai Zhao}, \bibinfo{person}{Cody Rivera},
  \bibinfo{person}{Megan~Hickman Fulp}, \bibinfo{person}{Robert Underwood},
  \bibinfo{person}{Sian Jin}, \bibinfo{person}{Xin Liang}, \bibinfo{person}{Jon
  Calhoun}, \bibinfo{person}{Dingwen Tao}, {and} \bibinfo{person}{Franck
  Cappello}.} \bibinfo{year}{2020}\natexlab{}.
\newblock \showarticletitle{cuSZ: An Efficient GPU-Based Error-Bounded Lossy
  Compression Framework for Scientific Data}.
\newblock  (\bibinfo{year}{2020}), \bibinfo{pages}{3--15}.
\newblock


\bibitem[\protect\citeauthoryear{Wallace}{Wallace}{1992}]%
        {wallace1992jpeg}
\bibfield{author}{\bibinfo{person}{Gregory~K Wallace}.}
  \bibinfo{year}{1992}\natexlab{}.
\newblock \showarticletitle{The {JPEG} still picture compression standard}.
\newblock \bibinfo{journal}{\emph{IEEE Transactions on Consumer Electronics}}
  \bibinfo{volume}{38}, \bibinfo{number}{1} (\bibinfo{year}{1992}),
  \bibinfo{pages}{xviii--xxxiv}.
\newblock


\bibitem[\protect\citeauthoryear{Wang, Wang, and Yeung}{Wang
  et~al\mbox{.}}{2015}]%
        {wang2015collaborative}
\bibfield{author}{\bibinfo{person}{Hao Wang}, \bibinfo{person}{Naiyan Wang},
  {and} \bibinfo{person}{Dit-Yan Yeung}.} \bibinfo{year}{2015}\natexlab{}.
\newblock \showarticletitle{Collaborative deep learning for recommender
  systems}. In \bibinfo{booktitle}{\emph{Proceedings of the 21th ACM SIGKDD
  international conference on knowledge discovery and data mining}}. ACM,
  \bibinfo{pages}{1235--1244}.
\newblock


\bibitem[\protect\citeauthoryear{Wang, Ye, Zhao, Wu, Li, Song, Xu, and
  Kraska}{Wang et~al\mbox{.}}{2018}]%
        {wang2018superneurons}
\bibfield{author}{\bibinfo{person}{Linnan Wang}, \bibinfo{person}{Jinmian Ye},
  \bibinfo{person}{Yiyang Zhao}, \bibinfo{person}{Wei Wu}, \bibinfo{person}{Ang
  Li}, \bibinfo{person}{Shuaiwen~Leon Song}, \bibinfo{person}{Zenglin Xu},
  {and} \bibinfo{person}{Tim Kraska}.} \bibinfo{year}{2018}\natexlab{}.
\newblock \showarticletitle{Superneurons: dynamic GPU memory management for
  training deep neural networks}. In \bibinfo{booktitle}{\emph{Proceedings of
  the 23rd ACM SIGPLAN Symposium on Principles and Practice of Parallel
  Programming}}. ACM, \bibinfo{pages}{41--53}.
\newblock


\bibitem[\protect\citeauthoryear{Wozniak, Jain, Balaprakash, Ozik, Collier,
  Bauer, Xia, Brettin, Stevens, Mohd-Yusof, et~al\mbox{.}}{Wozniak
  et~al\mbox{.}}{2018}]%
        {wozniak2018candle}
\bibfield{author}{\bibinfo{person}{Justin~M Wozniak}, \bibinfo{person}{Rajeev
  Jain}, \bibinfo{person}{Prasanna Balaprakash}, \bibinfo{person}{Jonathan
  Ozik}, \bibinfo{person}{Nicholson~T Collier}, \bibinfo{person}{John Bauer},
  \bibinfo{person}{Fangfang Xia}, \bibinfo{person}{Thomas Brettin},
  \bibinfo{person}{Rick Stevens}, \bibinfo{person}{Jamaludin Mohd-Yusof},
  {et~al\mbox{.}}} \bibinfo{year}{2018}\natexlab{}.
\newblock \showarticletitle{CANDLE/Supervisor: A Workflow Framework for Machine
  Learning Applied to Cancer Research}.
\newblock \bibinfo{journal}{\emph{BMC Bioinformatics}} \bibinfo{volume}{19},
  \bibinfo{number}{18} (\bibinfo{year}{2018}), \bibinfo{pages}{491}.
\newblock


\bibitem[\protect\citeauthoryear{You, Gitman, and Ginsburg}{You
  et~al\mbox{.}}{2017}]%
        {you2017scaling}
\bibfield{author}{\bibinfo{person}{Yang You}, \bibinfo{person}{Igor Gitman},
  {and} \bibinfo{person}{Boris Ginsburg}.} \bibinfo{year}{2017}\natexlab{}.
\newblock \showarticletitle{Scaling sgd batch size to 32k for imagenet
  training}.
\newblock \bibinfo{journal}{\emph{arXiv preprint arXiv:1708.03888}}
  \bibinfo{volume}{6} (\bibinfo{year}{2017}).
\newblock


\bibitem[\protect\citeauthoryear{Young, Hazarika, Poria, and Cambria}{Young
  et~al\mbox{.}}{2018}]%
        {young2018recent}
\bibfield{author}{\bibinfo{person}{Tom Young}, \bibinfo{person}{Devamanyu
  Hazarika}, \bibinfo{person}{Soujanya Poria}, {and} \bibinfo{person}{Erik
  Cambria}.} \bibinfo{year}{2018}\natexlab{}.
\newblock \showarticletitle{Recent trends in deep learning based natural
  language processing}.
\newblock \bibinfo{journal}{\emph{ieee Computational intelligenCe magazine}}
  \bibinfo{volume}{13}, \bibinfo{number}{3} (\bibinfo{year}{2018}),
  \bibinfo{pages}{55--75}.
\newblock


\bibitem[\protect\citeauthoryear{Zstandard}{Zstandard}{2020}]%
        {zstd}
\bibfield{author}{\bibinfo{person}{Zstandard}.}
  \bibinfo{year}{2020}\natexlab{}.
\newblock \bibinfo{howpublished}{\url{http://facebook.github.io/zstd/}}.
\newblock


\end{thebibliography}

\end{document}